\documentclass[useAMS,usenatbib]{mn2e}

\usepackage{epsfig}

\newcommand{\etal}{{et al.}}
\newcommand{\eg}{{e.g.,}}
\newcommand{\lya}{\ifmmode {\rm Ly\alpha}\else{\rm Ly$\alpha$}\fi}
\newcommand{\hzrgs}{HzRGs}
\newcommand{\hzrg}{HzRG}
\newcommand{\LFIR}{\ifmmode {\,L_{\rm FIR}}\else ${\,L_{\rm FIR}}$\fi}
\newcommand{\LUV}{\ifmmode {\rm \,L_{UV}}\else ${\rm \,L_{UV}}$\fi}
\newcommand{\Llya}{\ifmmode {\rm \,L_{Ly--\alpha}}\else ${\rm \,L_{Ly--\alpha}}$\fi}
\newcommand{\OmM}{\ifmmode {\Omega_{\rm M}}\else $\Omega_{\rm M}$\fi}
\newcommand{\OmL}{\ifmmode {\Omega_{\Lambda}}\else $\Omega_{\Lambda}$\fi}
\newcommand{\ph}{\ifmmode {h_{65}^{-1}}\else $h_{65}^{-1}$\fi}
\newcommand{\psqh}{\ifmmode {h_{65}^{-2}}\else $h_{65}^{-2}$\fi}

\newcommand{\degree}{\ifmmode {^{\,\circ}} \else {$^{\,\circ}$}\fi}
\newcommand{\Lsun}{\ifmmode {\rm\,L_\odot}\else ${\rm\,L_\odot}$\fi}
\newcommand{\Msun}{\ifmmode {\rm\,M_\odot} \else ${\rm\,M_\odot}$\fi}
\newcommand{\Zsun}{\ifmmode {\rm\,Z_\odot} \else ${\rm\,Z_\odot}$\fi}
\newcommand{\kmps}{\ifmmode {\rm\,km~s^{-1}} \else ${\rm\,km\,s^{-1}}$\fi}
\newcommand{\kpc}{{\rm\,kpc}} 
\newcommand{\ergps}{\ifmmode {\rm\,erg\,s^{-1}} \else {${\rm\,erg\,s^{-1}}$}\fi}
\newcommand{\ergpspcm}{\ifmmode {\rm\,erg\,s^{-1}\,cm^{-2}} \else {${\rm\,erg\,s^{-1}\,cm^{-2}}$}\fi}
\newcommand{\surfbr}{\ifmmode {\rm\,erg\,s^{-1}\,cm^{-2}\,arcsec^{-2}} \else {${\rm\,erg\
,s^{-1}\,cm^{-2}\,arcsec^{-2}}$}\fi}
\newcommand{\Msunpyr}{\ifmmode {\rm\,M_\odot\,yr^{-1}} \else {${\rm\,M_\odot\,yr^{-1}}$}\fi}
\newcommand{\pyr}{\ifmmode {\rm\,yr^{-1}} \else {${\rm\,yr^{-1}}$}\fi}
\newcommand{\psec}{\ifmmode {\rm\,s^{-1}} \else {${\rm\,s^{-1}}$}\fi}

\newcommand{\nSMMdetzR}{12} 
\newcommand{\nSMMobsz}{24} 
\newcommand{\nSMMobstotAR}{69}
\newcommand{\nSampleAR}{67} 

\newcommand{\aap}{A\&A}
\newcommand{\aaps}{A\&AS}
\newcommand{\aj}{AJ}
\newcommand{\apj}{ApJ}
\newcommand{\apjl}{ApJ}
\newcommand{\apjs}{ApJ}
\newcommand{\araa}{ARA\&A}
\newcommand{\mnras}{MNRAS}
\newcommand{\nat}{Nature}

\newcommand{\hi}{H\,{\sc i}}
\newcommand{\civ}{C\,{\sc iv}}
\newcommand{\nv}{N\,{\sc v}}
\newcommand{\oiii}{O\,{\sc iii}}

\newcommand{\lt}{$<$}

\title[Dust and star formation in distant radio galaxies]{Dust and
star formation in distant radio galaxies}

\author[M. Reuland, H. R\"ottgering, W. van Breugel, and C. De Breuck]{Michiel Reuland,$^{1,2,3}$\thanks{E-mail:
reuland@strw.leidenuniv.nl} Huub R\"ottgering,$^{1}$ Wil van
Breugel,$^{2}$ and Carlos De Breuck$^{4}$\\\\
$^{1}$Leiden Observatory, P.O. Box 9513, 2300 RA Leiden, The Netherlands\\
$^{2}$Institute of Geophysics and Planetary Physics, 
L--413 Lawrence Livermore National Laboratory,  
Livermore, CA 94550, USA\\
$^{3}$Department of Physics, UC Davis, 1 Shields Avenue, Davis, CA 95616, U.S.A.\\
$^{4}$Institut d'Astrophysique de Paris, CNRS, 98bis Boulevard Arago, F--75014 Paris, France}

\begin{document}

\date{Accepted XXXX December XX. Received XXXX December XX; in original form XXXX December XX}

\pagerange{\pageref{firstpage}--\pageref{lastpage}} \pubyear{XXXX}

\maketitle

\label{firstpage}

\begin{abstract}
We present the results of an observing program with the SCUBA
bolometer array to measure the submillimetre (submm) dust continuum
emission of \nSMMobsz\ distant ($z > 1$) radio galaxies.  We detected
submm emission in \nSMMdetzR\ galaxies with S/N $> 3$, including 9
detections at $z > 3$.  When added to previous published results these
data almost triple the number of radio galaxies with $z > 3$ detected
in the submm and yield a sample of \nSMMobstotAR\ observed radio
galaxies over the redshift range $z$ = 1--5. We find that the range
in rest-frame far-infrared luminosities is about a factor of 10. We
have investigated the origin of this dispersion, correlating the
luminosities with radio source power, size, spectral index, $K$-band
magnitude and \lya\ luminosity.  No strong correlations are apparent
in the combined data set.  We confirm and strengthen the result from
previous submm observations of radio galaxies that the detection rate
is a strong function of redshift. We compare the redshift dependence
of the submm properties of radio galaxies with those of quasars and
find that for both classes of objects the observed submm flux density
increases with redshift to $z \approx 4$, beyond which, for the
galaxies, we find tentative evidence for a decline.  We find evidence
for an anti-correlation between submm luminosity and UV polarisation
fraction, for a subsample of 13 radio galaxies, indicating that
starbursts are the dominant source of heating for dust in radio
galaxies.

\end{abstract}


\section{Introduction}

There is strong evidence that powerful high redshift radio galaxies
(\hzrgs; $z > 2$) are the progenitors of the brightest cluster
ellipticals seen today.  \hzrgs\ are the infrared brightest and
presumably the most massive galaxies at any epoch \citep{DeBreuck02aj}
and host actively-accreting super massive black holes with masses of
order $10^{9}\,\Msun$ \citep{Lacy01apj,Dunlop03mnras}.  Therefore,
they are key objects for studying the formation and evolution of
massive galaxies and super-massive black holes.

\hzrgs\ are likely to be in an important phase of their formation
process for several reasons: They have large reservoirs of gas from
which they could be forming, as shown by spectacular ($>$ 100 kpc)
luminous \lya\ haloes
\citep[\eg][]{McCarthy93araa,vanOjik96aa,Reuland03apj} and widespread
\hi\ absorption features in the \lya\ profiles \citep{vanOjik97aa}.
Their rest-frame UV morphologies are characterized by clumpy
structures, similar to the Lyman-break galaxies at $z \sim 3$ , that
will merge with the central galaxy on dynamical time-scales of
$10^8$\,yrs \citep{Pentericci98apj,Pentericci99aa}.  In the case of
4C~41.17 there is direct evidence for massive star formation (up to $\sim
1500\,\Msunpyr$ after correction for extinction) based on stellar
absorption--lines \citep{Dey97apj}.  Finally, mm-interferometry studies
of CO line and continuum emission for three $z > 3$ HzRGs have shown
that the star formation occurs galaxy wide over distances up to
30\,kpc \citep{Papadopoulos00apj,DeBreuck03aa} and there is even
evidence for star formation on scales of 250\kpc\
\citep{Stevens03nat}.  Together this suggests that we are observing
not merely scaled up versions of local ultraluminous infrared
galaxies (ULIRGs) where the bursts are confined to the inner few kpc,
but wide-spread starbursts within which the galaxies are forming the
bulk of their eventual stellar populations.

\hzrgs\ are an important sample for studying the star formation
history of the universe because their selection is based on long
wavelength radio emission whose propagation is not affected by the
presence of dust.  Dust is expected to play a significant role in star
forming regions, absorbing UV/optical radiation from the starburst and
reradiating it at far-infrared (FIR) wavelengths
\citep{SandersMirabel96araa}.  Optical searches for distant galaxies
\citep[\eg\ using the Lyman-break technique;][]{Steidel96apj,
Steidel99apj,Ouchi01apj} are thus likely to be biased against dusty
objects.  Finding distant star forming galaxies through submillimetre
(submm; rest-frame FIR) emission \citep[\eg\
][]{Hughes98nat,Bertoldi02confproc,Chapman02amnras,Cowie02aj,Scott02mnras,Smail02mnras,Webb03aapj,Eales03mnras}
selects only the most obscured sources.  So far there has been little
overlap between the optical and submm selected star forming sources
\citep[selection on very red near-IR colours may prove more fruitful;
\eg\ ][]{Frayer04aj}. It remains unclear whether they are members of a
continuous population \citep[\eg ][]{AdelbergerSteidel00apj,Webb03bapj}
and arguments have been made that either could dominate the star
formation density at high redshift
\citep{Blain99mnras,AdelbergerSteidel00apj}.  Since selection at radio
wavelengths circumvents the aforementioned selection biases it could
help determine the relative contributions of obscured and unobscured
star formation to the star formation history of the universe.

\citet[][hereafter A01]{Archibald01mnras} have conducted the first
systematic submm survey to study the star formation history of radio
galaxies over a redshift interval of $0.7 < z < 4.4$. In their sample
of 47 galaxies, they found evidence for a considerable range in FIR
luminosities, a substantial increase in 850\,\micron\ detection rate
with redshift and that the average 850\,\micron\ luminosity rises at a
rate $(1 + z )^{3-4}$ out to $z \simeq 4$.  These results prompt the
following questions: Is the dispersion in FIR luminosities due to
differences in their star formation rates or dust contents? Does the
strong increase with redshift reflect an increase in star formation
rates or could it be related to changing dust properties? Does the FIR
luminosity keep on rising with redshift or does it level off and is
there a redshift cut-off? Are the inferred star formation rates
comparable to those derived from the optical/UV? Do the submm
properties of quasars (QSOs) and radio galaxies show similar trends or
do the two classes of objects evolve differently?
 
The submm findings from A01 were based on a limited number of
detections at high redshift ($z > 3$).  To put these results on a
statistically firmer footing and search for possible correlations with
other galaxy parameters more submm observations were required.  Here
we present such observations of all $z > 3$ HzRGs known at the
beginning of 2001 \citep[\eg][]{DeBreuck01aj} which had not been
observed in the submm. Adding these to the survey of A01 almost
triples the number of detections at high redshift, creating a sample
which is statistically significant over the full redshift range $z =$ 1 -- 5.

The structure of this paper is a follows: the sample selection,
observations and data analysis are described in Section 2. Results and
notes on some individual sources are presented in Section 3. Various
correlations with submm properties of \hzrgs\ are investigated in
Section 4 and described in detail in Section 5. Section 6 presents a
comparison between \hzrgs\ and QSOs.  We discuss and summarize our
conclusions in Section 7.  Throughout this paper, we adopt a flat
universe with $\OmM = 0.3$, $\OmL = 0.7$, and $H_{0} = 65 \kmps\,\rm
Mpc^{-1}$. Using this cosmology the look-back time at $z \sim 2.5$
(the median redshift of our sample) is 11.7\,\ph\,Gyr and a galaxy at
such a redshift must be less than 2.8\,\ph\, Gyr old.

\section{Sample Selection and Observations}

The observations presented here include submm observations of distant
radio galaxies.  The targets were selected from an increasing sample
of \hzrgs\ that is the result of an ongoing effort by our group and
others \citep[][de Vries \etal\ in preparation; Spinrad private
communication]{DeBreuck00aas,DeBreuck01aj} to find
distant radio galaxies based on Ultra Steep Radio Spectrum (USS;
i.e. red radio color) and near-IR identification selection criteria
\citep[for details see][]{DeBreuck01aj}.

We selected all \hzrgs\ known at the beginning of 2001 with redshifts
$z \ga 3$ and declination $\delta > -30\degree$ that did not have
prior submm observations.  Our aim was to observe a significant sample
of \hzrgs\ to complement the observations of A01 and, in particular,
to obtain better statistics at the highest redshifts. MG~2141+192 and
4C~60.07 were observed in both programs, because, at the time of
observation, their inclusion in the A01 sample was unknown to us.  The
850\,\micron\ results for B3~J2330+3927, 6C~J1908+722, and 4C~60.07
have been published previously as part of their CO imaging studies
\citep{Papadopoulos00apj,DeBreuck03aa}.  MRC~1138$-$262 was included
in the program because of its wealth of supporting data \citep[\eg\
][]{Pentericci97aa,Carilli02apj} and WN~J1115+5016 because it is one
of only two radio galaxies showing a broad absorbtion line (BAL)
system \citep{DeBreuck01aj}, the other BAL radio galaxy, 6C~J1908+722,
being a strong CO emitter \citep{Papadopoulos00apj}.  WN~J0528+6549 at
$z=1.210$ was observed because it was first thought to be at redshift
$z=3.120$ (actually belonging to another galaxy on the slit).

The coordinates, redshifts, largest angular sizes of the radio
sources, \lya\ fluxes, $K$-band magnitudes, and references for the
full sample observed in the submm are listed in Table
\ref{TableRadio}.

\begin{table*}
\begin{minipage}{16cm}
\caption{Redshifts, radio positions, largest angular sizes, \lya\
fluxes, $K$-band magnitudes and references to papers from which these
data were taken for all objects that were observed in our submm
program. The \lya\ fluxes are in units of $10^{-16}$\ergpspcm, and the
$K$-band magnitudes were measured in a 64\,kpc diameter aperture where
possible.  References: DB99,DB00a,DB00b,DB01,DB02,DB03a,DB03b=
\citet[][De Breuck \etal\ in
preparation]{DeBreuck99aa,DeBreuck00aas,DeBreuck00aa,DeBreuck01aj,DeBreuck02aj,DeBreuck03aa},
dV03 = de Vries \etal\ in preparation, Dey99 = \citet{Dey99amsproc},
Dri97 = \citet{Drinkwater97mnras}, ER96 = \citet{EalesRawlings96apj},
McC90 = \citet{McCarthy90aj}, McC96 = \citet{McCarthy96apj}, Kap98 =
\citet{Kapahi98apj}, Pap00 = \citet{Papadopoulos00apj}, Pen97 =
\citet{Pentericci97aa}, Ren98 = \citet{Rengelink98thesis}, R\"ot97 =
\citet{Rottgering97aa}, S99 = \citet{Stern99aj}, vB99 =
\citet{vanBreugel99apj}.
\label{TableRadio}} 
\begin{tabular}{lcrrrrrrrrrl}
\hline
Source & $z$ & \multicolumn{3}{c}{RA(J2000)} & \multicolumn{3}{c}{DEC (J2000)} & LAS &
$F_{\lya}$ & $K$ & References\\ 
& & h & m & s & \degree & $'$ & $''$ & $''$ & cgs & mag \\
\hline
    WNJ0528+6549 &    1.210 &    5 &  28 &  46.07 &  $ + $  65 &  49 &  57.3 &   1.9 & $-$ &  18.2   & DB00a, dV03 \\         
     MRC1138$-$262 &    2.156 &   11 &  40 &  48.25 &  $ - $  26 &  29 &  10.1 &  15.8 &  13.9 &  16.1     & R\"ot97, Pen97, DB00b \\ 
    WNJ1115+5016 &    2.550 &   11 &  15 &   6.87 &  $ + $  50 &  16 &  23.9 &   0.2 &   2.0 &  19.2     & DB00a, DB01, DB02 \\         
    WNJ0747+3654 &    2.992 &    7 &  47 &  29.38 &  $ + $  36 &  54 &  38.1 &   2.1 &   0.8 &  20.0     & DB00a, DB01, DB02 \\         
    WNJ0231+3600 &    3.079 &    2 &  31 &  11.48 &  $ + $  36 &   0 &  26.6 &  14.8 &   1.1 & $-$   & DB00a, DB01, DB02 \\         
    B3J2330+3927 &    3.086 &   23 &  30 &  24.91 &  $ + $  39 &  27 &  11.2 &   1.9 &   4.4 &  18.8     & DB03a \\         
    TNJ1112$-$2948 &    3.090 &   11 &  12 &  23.86 &  $ - $  29 &  48 &   6.2 &   9.1 &   2.9 & $-$   & DB00a, DB01 \\         
     MRC0316$-$257 &    3.130 &    3 &  18 &  12.06 &  $ - $  25 &  35 &   9.7 &   7.6 &   2.4 & $-$   & McC90, ER96, DB00b \\         
     PKS1354$-$17 &    3.150 &   13 &  47 &  96.03 &  $ - $  17 &  44 &  02.2 & $-$       &  $-$     &   $-$   & Dri97 \\  
   WNJ0617+5012 &    3.153 &    6 &  17 &  39.37 &  $ + $  50 &  12 &  54.7 &   3.4 &   0.8 &  19.7     & DB00b, DB01, DB02 \\         
     MRC0251$-$273 &    3.160 &    2 &  53 &  16.70 &  $ - $  27 &   9 &   9.6 &   3.9 & $-$ & $-$ & McC96, Kap98 \\         
    WNJ1123+3141 &    3.217 &   11 &  23 &  55.85 &  $ + $  31 &  41 &  26.1 &  25.8 &   6.2 &  17.5     & DB00a, DB01, DB02 \\         
    WNH1702+6042 &    3.223 &   17 &   3 &  36.23 &  $ + $  60 &  38 &  52.2 &  11.5 & $-$ & $-$ & Ren98\\         
    TNJ0205+2242 &    3.506 &    2 &   5 &  10.69 &  $ + $  22 &  42 &  50.3 &   2.7 & $-$ &  18.8   & DB00a, DB01, DB02\\         
    TNJ0121+1320 &    3.516 &    1 &  21 &  42.74 &  $ + $  13 &  20 &  58.3 &   0.3 & $-$ &  18.8   & DB00a, DB01, DB02\\         
      6C1908+722 &    3.532 &   19 &   8 &  23.70 &  $ + $  72 &  20 &  11.8 &  14.4 &  32.0 &  16.5     & Dey99, Pap00, DB01 \\         
    WNJ1911+6342 &    3.590 &   19 &  11 &  49.54 &  $ + $  63 &  42 &   9.6 &   1.8 &   1.4 &  18.6     & DB00a, DB01, DB02 \\         
      MG2141+192 &    3.592 &   21 &  44 &   7.50 &  $ + $  19 &  29 &  15.0 &   8.5 &   6.2 &  19.3     & S99\\         
    WNJ0346+3039 &    3.720 &    3 &  46 &  42.68 &  $ + $  30 &  39 &  49.3 &   0.4 & $-$ &  17.8   & DB00a, DB02, dV03 \\
         4C60.07 &    3.791 &    5 &  12 &  55.15 &  $ + $  60 &  30 &  51.0 &  16.0 &  10.1 &  19.3     & R\"ot97, Pap00, DB00b \\       
    TNJ2007$-$1316 &    3.830 &   20 &   7 &  53.23 &  $ - $  13 &  16 &  43.6 &   7.2 &   2.5 &  17.9     & DB00a, DB02, DB03b\\         
    TNJ1338$-$1942 &    4.100 &   13 &  38 &  26.06 &  $ - $  19 &  42 &  30.1 &   5.5 &  10.1 &  19.7     & DB99 \\         
    TNJ1123$-$2154 &    4.109 &   11 &  23 &  10.15 &  $ - $  21 &  54 &   5.3 &   0.8 &   0.2 &  20.4     & DB00a, DB01, DB02 \\         
    TNJ0924$-$2201 &    5.190 &    9 &  24 &  19.92 &  $ - $  22 &   1 &  41.5 &   1.2 &   0.4 &  19.9     & vB99 \\         
\hline
\end{tabular}
\end{minipage}
\end{table*}

\subsection{SCUBA photometry} 
The observations were carried out between October 1997 and January
2002 with the Submillimetre Common--User Bolometer Array
\citep[SCUBA;][]{Holland99mnras} at the 15\,m James Clerk Maxwell
Telescope (JCMT).  We observed at 450\,\micron\ and 850\,\micron\
wavelengths resulting in beam sizes of 7.5\arcsec\ and 14.7\arcsec\
respectively. We employed the 9-point jiggle photometry mode, which
samples a 3 $\times$ 3 grid with 2\arcsec\ spacing between grid
points, while chopping 45\arcsec\ in azimuth at 7.8\,Hz.  Frequent
pointing checks were performed to ensure pointings better than
2\arcsec\ and reach optimal sensitivity.

Our original goal was to observe all sources down to 1\,mJy rms at
850\,\micron. This is a sensible limit, since at 850\,\micron\
confusion becomes a problem for sources weaker than 2\,mJy
\citep{Hughes98nat,Hogg01aj}, it is obtainable in 3\,hrs per source,
and it is matched to the survey by A01. However, because of scheduling
constraints, and because our priority was to obtain a large sample of
\hzrgs\ with 850\,\micron\ detections, this limit was not always
reached.  Rather, the next target was observed as soon as an apparent
5$\sigma$ detection had been obtained at 850\,\micron.

The atmospheric optical depths $\tau_{850}$, $\tau_{450}$ were
calculated using the empirical CSO--tau correlations given by
\citet{Archibald02mnras}, unless the values obtained through skydips
disagreed strongly, in which case those were used instead.  The
optical depth $\tau_{850}$ varied between 0.14 and 0.38 with an
average value of 0.26.  The data were clipped at the 4$\sigma$ level
to ensure accurate determination of the sky level, flat--fielded,
corrected for extinction, sky noise was removed after which they were
co--added and clipped at the 2.5$\sigma$ level using the Scuba User
Reduction Facility software package \citep[SURF;][]{Jenness98adass},
following standard procedures outlined in the SCUBA Photometry
Cookbook\footnote{The SCUBA Photometry Cookbook is available at
http://www.starlink.rl.ac.uk/star/docs/sc10.htx/sc10.html}.  The
concatenated data were checked for internal consistency using a
Kolmogorov--Smirnov (K--S) test and severely deviating measurements
(if any) were removed. Finally, flux calibration was performed using
HLTAU, OH231.8 and CRL618 as photometric calibrators. The typical
photometric uncertainty for our program is of order 10--15 per cent,
as estimated from our results on three sources (TN~J0121+1320,
TN~J1338$-$1942, and MG~2141+192) that were observed at two separate
instances each. This photometric accuracy is consistent with an
estimated 10 per cent systematic uncertainty in the 850\,\micron\ flux
density scale \citep[see \eg][]{Papadopoulos00apj,Jenness01adass}.
Given that \hzrgs\ appear to be located in submm overdense regions
\citep[\eg\ ][]{Stevens03nat} flux may have been lost due to chopping
onto a nearby galaxy. However this very unlikely to have affected more
than a few sources.

Following \citet{Omont01aa}, Table \ref{Table850}
includes a column indicating the quality of the observation. Good
quality data is indicated by an `A', whereas poor quality is indicated
by a `B'. Poor quality reflects bad atmospheric conditions (e.g. large
seeing), short integration time ($< 2$ sets of 50 integrations
each), or poor internal consistency as shown by the K--S test (i.e. the
measurements were not consistent, but it was impossible to determine
which were the outliers. In such cases the average of all measurements
was used).

\subsection{Potential contamination of the thermal submillimetre flux}
\label{Contaminated}
Because all our objects are powerful radio galaxies, it is important
to estimate any synchrotron contribution to the observed submm band.
We used flux densities from the WENSS
\citep[325\,MHz][]{Rengelink97aa}, Texas
\citep[365\,MHz;][]{Douglas96aj} and NVSS
\citep[1.4\,GHz;][]{Condon98aj} surveys to extrapolate to 350\,GHz
(850\,\micron) frequencies using a power law.  For 53W069, we
extrapolated from the 600\,MHz and 1.4\,GHz values in
\citet[]{Waddington00mnras}. We find that the synchrotron contribution
at 850\,\micron\ is negligible for most galaxies in our sample. Only
for some objects from A01 (and PKS~1354$-$17) would this require
corrections larger than the 1$\sigma$ uncertainties in the
850\,\micron\ measurements.

Synchrotron spectra often steepen at high frequencies \citep[\eg\
A01;][]{Athreya97mnras,Andreani02aa,Sohn03aa}, and linear
extrapolation should be considered an upper limit to the synchrotron
contribution. A01 performed parabolic fits to account for the
curvature of the radio spectrum. Using the midpoint between linear and
parabolic fits they find corrections larger than 1.5\,mJy to the
850\,\micron\ flux densities in only 6 cases.  One could argue that
parabolic fits are more appropriate, in which case all corrections
would be negligible.

Note that the radio measurements reflect the spatially integrated flux
densities of the sources.  The radio cores have flatter spectra than
the lobes and could dominate at higher frequencies. However, they are
usually faint and for USS sources even the radio cores tend to have
steep spectra \citep[\eg\ A01;][]{Athreya97mnras} indicating that
contamination by the core is likely to be negligible as well.

Given these uncertainties, extrapolation from the radio regime to
submm wavelengths is uncertain and is likely to result in an
overestimate of the non-thermal contribution due to steepening of the
radio spectrum.  This is demonstrated for the case of B3~J2330+3927,
for which \citet{DeBreuck03aa} estimate a non-thermal contribution of
$\sim1.3$\,mJy at 113\,GHz but measured a flux density $< 0.3$\,mJy
that seems to be of thermal origin.  Therefore we do not correct for a
contribution to the submm continuum from the non-thermal radio
emission, except for a few sources discussed below.  For these reasons
\citep[and following][]{Willott02mnras} we have chosen also to use
uncorrected fluxes from A01 in the remainder of this paper.

There are two exceptions that we exclude from our final sample.
Following A01, we reject B2~0902+34, as this source has a bright
flat-spectrum radio core which could dominate the submm emission. We
also reject the flat spectrum radio source PKS~1354$-$17. It is
significantly brighter than any of the other sources at 850\,\micron\
($S_{850} = 20.5 \pm 2.6$\,mJy), but linear extrapolation from the
radio regime shows that the non-thermal contribution at 850\,\micron\
could be as large as $40$\,mJy and could easily account for all of the
submm signal.

Gravitational lensing may be important in some cases
\citep{Lacy99confproc}, resulting in enhanced submm fluxes. However,
recent estimates \citep[\eg][]{Chapman02bmnras,Dunlop02astroph} show
that this is limited to a small but significant fraction (3--5 per cent of
sources with $S_{\rm 850} > 10$\,mJy may have been boosted by a factor
$\ga 2$) and that for most objects there is no evidence for strong
gravitational lensing. Corrections for lensing must be made on a
case-to-case basis and are strongly model dependent.  Since these
corrections are likely to be small, they have not been attempted for
the present sample.
 
\subsection{Dust template}
\noindent As has been noted by many authors \citep[\eg\
A01;][]{Hughes97mnras}, choosing the dust template is an important
step in inferring the bolometric far infra-red luminosities (\LFIR),
star formation rates (SFR) and dust masses ($M_{\rm d}$).  A
complication is that the appropriate dust template may change over
redshift due to changing dust properties with the evolutionary states
of the galaxies.

Throughout this paper we adopt single temperature, optically thin
greybody emission for two sets of emissivity index $\beta$ and
temperature $T_{\rm d}$ \citep[see ][for possible
concerns]{Dunne01mnras,Dupac03aa} as the functional parametrization
for thermal dust emission from high redshift sources. We choose $\beta
= 1.5$ and $T_{\rm d} = 40$\,K for comparison with other papers
\citep[\eg\ A01; and see][]{Dunne00mnras, Eales03mnras}, and also
briefly investigate the effects of assuming $\beta = 2.0$ and $T_{\rm
d} = 40$\,K as seems reasonable for some hyperluminous IR galaxies
\citep[$T_{\rm d} = 35$\,K; HyLIRGs][]{Farrah02mnras} and $z > 4$
quasars \citep[$T_{\rm d} = $
40--50\,K;][]{PriddeyMcMahon01mnras,Willott02mnras}.  Measuring the
value of $\beta$ for \hzrgs\ directly would require observations at
many more rest-frame FIR wavelenghts than presented here. Increasing
$\beta$ or the dust temperature decreases the inferred $\LFIR$ of high
redshift sources relative to lower redshift sources for a given flux
density.

The fraction of absorbed UV/optical light, $\delta_{\rm SB}$, and
possible departures from the prototype Salpeter initial mass function,
parametrized with $\delta_{\rm IMF}$, are other uncertain
factors. Generally accepted approximations \citep[see
\eg][]{Papadopoulos00apj,Omont01aa,DeBreuck03aa} for the dust mass,
inferred FIR luminosity and star formation rate are respectively:
$$M_{\rm d} = {{S_{\rm obs}D_{\rm L}^{2}} \over {(1+z)\kappa_{\rm
d}(\nu_{\rm rest}) B(\nu_{\rm rest},T_{\rm d})}},$$
$$\LFIR = 4\pi M_{\rm d} \int_{0}^{\infty} \kappa_{\rm d}(\nu)
B(\nu,T_{\rm d}) d\nu = $$
$${8\pi h \over c^{2}} {\kappa_{\rm d}(\nu) \over \nu^{\beta}} \left(
\frac{kT}{h} \right)^{\beta+4} \Gamma(\beta +4) \zeta(\beta+4) M_{\rm
d},$$ and
$$SFR = \delta_{\rm IMF} \delta_{\rm SB} (\LFIR/10^{10}\Lsun)
\Msunpyr,$$ with $\kappa_{\rm d}(\nu) \propto \nu^{\beta}$
the frequency dependent mass absorption coefficient which modifies the
Planck function, $B(\nu,T_{\rm d})$, to describe the
isothermal greybody emission from dust grains, $\Gamma$ the Gamma
function, $\zeta$ the Riemann Zeta function, $D_{\rm L}$ the luminosity
distance and $S_{\rm obs}$ the observed flux density. The mass
absorption coefficient is poorly constrained
\citep[\eg][]{Chini86aa,Downes92apj,DeBreuck03aa,James02mnras} and we
conform to the intermediate value of $\kappa_{\rm d}(375\rm GHz) =
0.15\, \rm m^{2}\,kg^{-1}$ chosen by A01.

\section{Observational Results}\label{ObsResults}

\nSMMobsz\ radio sources were observed. The results of the
observations, the inferred rest-frame 850\,\micron\ luminosities
$L_{\rm 850}$, total far-IR luminosities \LFIR, and radio
luminosities $L_{\rm 3 GHz}$ are summarized in Table \ref{Table850}.
\nSMMdetzR\ of the \hzrgs\ are detected at $>$3$\sigma$ significance
at 850\,\micron. The median rms flux density of the observations is
$\sigma_{850} = 1.5$\,mJy with an interquartile range of $0.8$\,mJy.
Only B3~J2330+3927 may have been detected at 450\,\micron\ at a
$2\sigma$ level ($S_{450} = 49.1 \pm 17.7$\,mJy).

\begin{table*}
\begin{minipage}{18cm}
\caption{Observed 850\,\micron\ and 450\,\micron\ submm flux densities
$S_{\rm 850\,\micron}$ and $S_{\rm 450\,\micron}$ with their standard
errors for the radio sources in the program. The total duration of the
observations, $N_{\rm int}$ is given in sets of 50
integrations. 3$\sigma$ upper limits to the 850\,\micron\ flux are
shown for sources whose S/N is below 3. Only B3~J2330+3927 may have
been detected at 2$\sigma$ at 450\,\micron. Logarithms of inferred
rest-frame 850\,\micron\ luminosities $L_{\rm 850}$, far-IR
luminosities, \LFIR\, and radio luminosities $L_{\rm 3 GHz}$ are shown
for the dust template with $\beta = 1.5$, $T_{\rm d} = 40$\,K and a
flat universe with $\OmM = 0.3$, $\OmL = 0.7$, and $H_{0} = 65
\kmps\,\rm Mpc^{-1}$.\label{Table850}}
\begin{tabular}{lcrcrccrrrc}
\hline
Source & $z$ & $N_{\rm int}$ &
$S_{850}$$^{a}$ & S/N &
Quality & 3$\sigma$ lim. & $S_{450}$ &  $L_{850}$\phantom{aa} & $L_{\rm FIR}$ & $L_{3 \rm GHz}$\\ 
& & $\times$ 50 & mJy & & & mJy & mJy & W\,Hz$^{-1}$\,sr$^{-1}$\hspace{-0.5cm} & \Lsun\ &  W\,Hz$^{-1}$\,sr$^{-1}$ \\
\hline
   WNJ0528+6549   &    1.210 & 4+4 &   $-$1.9 $ \pm$   1.3\phantom{$-$} &  $-$1.4 & A & $<$3.9    &    2 $ \pm$  29 &   $<$23.05\phantom{$^{b}$}  &$<$12.63\phantom{$^{b}$}  &   24.59  \\              
   MRC1138$-$262  &    2.156 & 2 &     12.8 $ \pm$   3.3$^{b}$ &     3.9 & B &                        &  -65 $ \pm$  134&       23.26$^{b}$              & 12.83$^{b}$ &   27.15 \\
   WNJ1115+5016   &    2.550 & 4 &      3.0 $ \pm$   1.3 &     2.3 & A & $<$6.9                     &  -20 $ \pm$  11 &  $<$23.31\phantom{$^{b}$}   & $<$12.90\phantom{$^{b}$} &   25.95  \\              
   WNJ0747+3654   &    2.990 & 6 &      4.8 $ \pm$   1.1 &     4.5 & A &                              &   18 $ \pm$  15 &        23.15\phantom{$^{b}$}   & 12.73\phantom{$^{b}$} &   26.22  \\              
   WNJ0231+3600   &    3.080 & 7 &      5.9 $ \pm$   1.6 &     3.7 & B &                              &  -29 $ \pm$  22 &        23.23\phantom{$^{b}$}   & 12.81\phantom{$^{b}$} &   26.27  \\              
   B3J2330+3927   &    3.086 & 3 &     14.1 $ \pm$   1.7$^{c}$ &     8.5 & A &                        &   49 $ \pm$  18 &       23.61\phantom{$^{b}$}    & 13.19\phantom{$^{b}$}  &   26.56  \\              
   TNJ1112$-$2948 &    3.090 & 5 &      5.8 $ \pm$   1.1 &     5.1 & A &                              &   15 $ \pm$   9 &        23.23\phantom{$^{b}$}   & 12.81\phantom{$^{b}$} &   26.66  \\              
   MRC0316$-$257  &    3.130 & 2 &      0.6 $ \pm$   2.7 &     0.2 & B & $<$8.8                     &    5 $ \pm$  48 & $<$23.41\phantom{$^{b}$}    & $<$12.99\phantom{$^{b}$} &   27.22  \\              
     PKS1354$-$17 &    3.150 & 2 &     20.5 $ \pm$   2.6$^{d}$ &     8.0 & B &                        &  -47 $ \pm$  77 &       23.77\phantom{$^{b}$}    & $-$\phantom{$^{b}$}  &   27.71 \\              
   WNJ0617+5012   &    3.153 & 6+6 &      1.0 $ \pm$   0.7 &     1.3 & B & $<$3.2                 &    3 $ \pm$  16 &  $<$22.96\phantom{$^{b}$}   & $<$12.55\phantom{$^{b}$} &   26.10  \\              
   MRC0251$-$273  &    3.160 & 2 &      0.6 $ \pm$   2.8 &     0.2 & A & $<$8.9                     &  -54 $ \pm$  91 & $<$23.41\phantom{$^{b}$}    & $<$12.99\phantom{$^{b}$} &   27.00  \\              
   WNJ1123+3141   &    3.220 & 8 &      4.9 $ \pm$   1.2 &     4.1 & A &                              &    4 $ \pm$  14 &        23.15\phantom{$^{b}$}   & 12.73\phantom{$^{b}$} &   26.60  \\              
   WNH1702+6042   &    3.223 & 1 &   $-$0.4 $ \pm$   3.6\phantom{$-$} &  $-$0.1 & B & $<$10.8       &  -73 $ \pm$  91 & $<$23.49\phantom{$^{b}$}    & $<$13.07\phantom{$^{b}$} &   26.40  \\              
   TNJ0205+2242   &    3.506 & 6 &      1.3 $ \pm$   1.3 &     1.0 & A & $<$5.2                     &   27 $ \pm$  23 &  $<$23.17\phantom{$^{b}$}   & $<$12.75\phantom{$^{b}$} &   26.58  \\              
   TNJ0121+1320   &    3.517 & 7+4 &      7.5 $ \pm$   1.0 &     7.6 & A &                          &    4 $ \pm$  16 &        23.33\phantom{$^{b}$}   & 12.91\phantom{$^{b}$} &   26.55  \\              
    6CJ1908+722   &    3.532 &  6 &     10.8 $ \pm$   1.2$^{c}$ &     9.0 & A &                       &  33 $ \pm$   17 &        23.49\phantom{$^{b}$}   & 13.07\phantom{$^{b}$} &   27.25 \\
   WNJ1911+6342   &    3.590 & 2 &      1.3 $ \pm$   3.6 &     0.4 & B & $<$11.9                    &  -38 $ \pm$  50 & $<$23.53\phantom{$^{b}$}    & $<$13.11\phantom{$^{b}$} &   26.26  \\              
   MG2141+192     &    3.592 & 7+5 & \phantom{$^{b,e}$}2.3 $ \pm$ 0.9$^{b,e}$ & 2.6 & A & $<$5.0  &   12 $ \pm$  13 &   22.96$^{b}$                  & $<$12.55$^{b}$ &   27.30 \\ 
   WNJ0346+3039   &    3.720 & 4 &   $-$0.5 $ \pm$   1.3\phantom{$-$} &  $-$0.4 & A & $<$3.8        &   -5 $ \pm$  12 &  $<$23.02\phantom{$^{b}$}   & $<$12.61\phantom{$^{b}$} &   26.43  \\  
        4C60.07   &    3.791 & 5 & 	 \phantom{$^{b,c}$}11.5 $ \pm$   1.5$^{b,c,e}$&   7.6 &  A &  &   10 $ \pm$  13     &   23.61$^{b}$              & 13.19$^{b}$ &   27.15 \\ 
   TNJ2007$-$1316 &    3.830 & 5 &      5.8 $ \pm$   1.5 &     4.0 & A &                              &    4 $ \pm$  45 &       23.21\phantom{$^{b}$}    & 12.79\phantom{$^{b}$} &   26.98  \\              
   TNJ1338$-$1942 &    4.100 & 4+7 &      6.9 $ \pm$   1.1 &     6.2 & A &                          &  -36 $ \pm$  32 &        23.29\phantom{$^{b}$}   & 12.87\phantom{$^{b}$} &   27.05  \\              
   TNJ1123$-$2154 &    4.109 & 2 &      1.5 $ \pm$   1.7 &     0.9 & A & $<$6.7                     &   -7 $ \pm$  11 &  $<$23.27\phantom{$^{b}$}   & $<$12.85\phantom{$^{b}$} &   26.76  \\              
   TNJ0924$-$2201 &    5.190 & 8+4 &   $-$0.7 $ \pm$   1.1\phantom{$-$} &  $-$0.7 & A &  $<$3.2 &    -0 $ \pm$  26 & $<$22.94\phantom{$^{b}$}    & $<$12.53\phantom{$^{b}$} &   27.24  \\              
\hline
\end{tabular}\\
$^{a}$This does not include the 10--15 per cent uncertainty in
absolute photometric calibration.\\
$^{b}$For the statistical analysis we use $S_{850} = 5.9 \pm
1.1$\,mJy, $S_{850} = 3.3 \pm 0.7$\,mJy and $S_{850} = 14.4 \pm
1.0$\,mJy for MRC~1138$-$262, MG~2141+192 and 4C~60.07,
respectively. $L_{850}$ and \LFIR\ were inferred using those values. See Section
\ref{Analysis} for details.\\
$^{c}$Data published in CO imaging studies by \citet{Papadopoulos00apj} and \citet{DeBreuck03aa}.\\
$^{d}$PKS~1354$-$17 is likely to be dominated by
non-thermal emission (c.f. Section \ref{Contaminated}).\\
$^{e}$Also part of the survey by A01.
\end{minipage}
\end{table*}

Particularly noteworthy is MG~2141+192. This source was detected by
A01 at the $4.8\sigma$-level at $S_{850} = 4.61 \pm 0.96$\,mJy, using
the narrow filterset whereas we observed $S_{850} = 2.16 \pm
1.10$\,mJy and $S_{850} = 2.45 \pm 1.58$\,mJy at two separate
instances with the wide filter set and did not detect the source.
Similarly 4C~60.07 has been detected in photometry mode at $S_{850} =
11.5 \pm 1.5, 17.1 \pm 1.3$ and in a jiggle map at $21.6 \pm 1.3$
\citep[A01;][]{Papadopoulos00apj,Stevens03nat}.  These measurements are
consistent to within 2--3$\sigma$ from the mean. However, naively,
they could also be interpreted as signs of submm variability.  A01 and
\citet{Willott02mnras} also found tentative evidence for variability
in the submm, but ascribed it to problems with sky subtraction for
data obtained with the single-element bolometer UKT14 versus SCUBA.
It is hard to see how widespread star formation could result in submm
variability on the time-scale of years. Significant changes in \LFIR\
might be easier to envisage as the result of UV variability often seen
in AGN and if the submm emission results from quasar heated
dust. However, even in this scenario changes in the UV are expected to
average over time in the observed submm regime, unless the FIR
emitting region is compact. While the typical scale sizes for UV
emission from the AGN are on parsec scales, the minimum extent of the
FIR emitting region must be about 1\,kpc to match the observed
luminosity and dust temperature \citep[\eg][]{Carilli01apj}.  Submm
variability, if real, is therefore hard to explain if reprocessing by
dust is the dominant mechanism for the FIR emission.  If,
alternatively, the FIR emission would be non-thermal emission from the
AGN, then the emission should be unresolved in contrast to the
observed extents of a few tens of kpc. Moreover, observations indicate
that AGN contribute at most 30 per cent of the FIR luminosity at
wavelengths longer than 50\,\micron, at least for HyLIRGs
\citep{RowanRobinson00mnras,Farrah02mnras}.

Possible explanations therefore may be that (i) for different chopping
angles and distances sometimes flux is accidentally lost due to
companion galaxies in the off-beam since the fields are overdense
\citep{Stevens03nat}, or (ii) that sometimes the AGN do in fact
contribute close to the maximum amount expected \citep[note that the
submm flux for 4C~60.07 is centrally concentrated; Fig. 2 in
][]{Stevens03nat}, even though the starbursts still dominate. (iii)
Finally of course there is still the possibility of pointing errors,
uncertainties in absolute flux calibration, and differences in
atmospheric transparency that could shift the effective bandpass by a
few GHz, which could make a difference due to the very steep slope of
the spectrum. 

\section{Analysis}\label{Analysis}

\noindent In the following we discuss a sample of \nSampleAR\ radio
galaxies (46/47 from A01, 23/24 from this paper, two sources were observed
in both samples). This excludes the flat-spectrum sources B2~0902+34
and PKS~1354$-$17 (see Section \ref{Contaminated}).  For the
statistical analysis we use the inverse variance weighted averages of
the measurements of MG~2141+192 ($<$$S_{\rm 850}$$> = 3.3 \pm 0.7$\,mJy)
and 4C~60.07 ($<$$S_{\rm 850}$$> = 14.4 \pm 1.0$\,mJy) and for
MRC~1138$-$262 we prefer the value of $S_{\rm 850} = 5.9 \pm 1.1$\,mJy
obtained by \citet{Stevens03nat} over our observation under adverse
conditions.  The median rms flux density for this entire sample is
$\sigma_{850} = 1.1$\,mJy with a 0.24\,mJy interquartile range.

Figure \ref{zS850} shows that the observed submm flux densities and
therefore the inferred luminosities (see Table \ref{Table850}) at $z >
3$ vary significantly from object to object. For the assumed dust
template we find a range from $\LFIR\ < 4 \times 10^{12}~\Lsun$ for
undetected targets to $\LFIR\ \sim 2 \times 10^{13}~\Lsun$ for
detected sources. There are several viable scenarios to explain this.
First, if all the warm dust is heated solely by young stars, then
\LFIR\ is linked to the SFR, implying that the SFR differs
significantly between objects.  Alternatively, there may be a range in
produced dust masses as substantial dust production may take more than
a billion years if low-mass stars are the principal contributors
\citep[\eg][]{Edmunds01mnras}.  In this case the range in \LFIR\ may
reflect a range in starburst ages. The recent detection of significant
amounts of dust in the local supernova remnant Cassiopeia A
\citep{Dunne03nat}, indicates that much faster evolving massive stars
may be at least as important in producing dust.  A considerable
contribution from high-redshift supernovae would significantly lower
the required time-scales for dust production and would favor a range in
SFRs to explain the range in \LFIR.

Despite the various uncertainties, the result that the far infrared
luminosities are high is robust.  This has important implications for
the starburst nature of these galaxies: inserting the values for
\LFIR\ that we find using either dust template confirms that \hzrgs\
are vigorously forming stars up to rates of a few 1000\Msunpyr\ and
have dust masses of a few times $10^{8}$\Msun. This is consistent with
the notion that they are in a critical phase of their formation,
forming the bulk of their stellar masses.

\subsection{Statistical analysis}

We have performed statistical tests to search for possible
correlations of the submm with other properties of the radio
galaxies. Specifically, we investigated whether there are correlations
with redshift (as reported by A01), radio luminosity (as indicator of
AGN contribution), largest angular size of the radio source (as
indicator of age), $K$-band magnitude (as indicator of stellar mass or
star formation rate), \lya\ flux (as a possible indicator of starburst
``fuel''), and UV polarisation (as indicator of the relative
contributions of starburst and AGN to the UV continuum). Below, we
discuss these in more detail.

Since the sample contains a large fraction (of order 50 per cent) of
non-detections (i.e. upper limits) we have conducted survival analysis
tests similar to A01. Survival analysis allows the mixing of
detections and upper limits (``censored data points'') in a
statistically meaningful way thereby preserving as much information as
possible. The results of the survival tests are summarized in Tables
\ref{SurvA_3} and \ref{SurvA_hz3}. Table \ref{SurvA_3} represents the
results for the entire sample, whereas Table \ref{SurvA_hz3} only
considers the 32 galaxies with redshifts $z > 2.5$.

\begin{table*}
\caption{Results of survival analyis for all 67 sources. The
significance $P$ is the probability of the variables not being
correlated according to each test. If all three tests yield $P \la$
5 per cent, then the variables are taken to be correlated. If only some
yield $P \la 5$ per cent then a correlation is regarded as possible, but
uncertain. Results are shown for both a 3$\sigma$ and 2$\sigma$
(between brackets) detection limit. \label{SurvA_3}}
\begin{tabular}{ccrrrrc}
\hline
Variable & & Percentage of &  & Significance ($P$) & & Correlation\\
Dependent & Independent & Data Censored
& Cox  & Kendall & Spearman & Present?\\
\hline
$S_{850\micron}$ & $z$                & 67\% (52)\%&  0\%  ( 0\%)&  0\% ( 0\%)&  0\% ( 0\%) & YES   (YES)   \\
$D_{\rm lin}$ & $L_{\rm 3\,GHz}$               &  0\% & 17\%  & 34\% & 32\%  & NO     \\
$L_{\rm 3\,GHz}$ & $z$                &  0\% &  1\%  &  2\% &  2\%  &  YES   \\
$L_{850\micron}$ & $L_{\rm 3\,GHz}$   & 67\% (52)\%&  1\%  ( 1\%)&  7\% ( 8\%)& 10\% ( 8\%) & MAYBE (MAYBE) \\
$L_{850\micron}$ & $z$                & 67\% (52)\%&  0\%  ( 0\%)&  0\% ( 0\%)&  0\% ( 0\%) & YES   (YES)   \\
$L_{850\micron}$ & $\alpha$           & 67\% (52)\%&  1\%  ( 1\%)&  1\% ( 0\%)&  1\% ( 0\%) & YES   (YES)   \\
$L_{850\micron}$ & $D_{\rm lin}$      & 67\% (52)\%& 56\%  (22\%)& 79\% (89\%)& 96\% (78\%) & NO    (NO)   \\
$\alpha$    & $L_{\rm 3\,GHz}$        &  0\%  & 36\%  & 20\% & 19\%  & NO        \\
$\alpha$    &  $z$                    &  0\% &  0\%  &  0\% &  0\%  & YES     \\
\hline
\end{tabular}
\end{table*}

\begin{table*}
\begin{minipage}{16cm}
\caption{Similar to Table \ref{SurvA_3}. Results of survival
analysis for 32 sources at redshifts  $z > 2.5$. \label{SurvA_hz3}} 
\begin{tabular}{ccrrrrc}
\hline
Variable & & Percentage of &  & Significance (P) & & Correlation\\
Dependent & Independent & Data Censored
& Cox$^{a}$
 & Kendall & Spearman & Present?\\
\hline
$S_{850\micron}$ & $z$                & 53\% (34\%)& 95\% (81\%)& 17\% (25\%)& 34\% (38\%) & NO    (NO) \\
$D_{\rm lin}$ & $L_{\rm 3\,GHz}$      &  0\% & 66\% & 8\% & 9\%  & NO   \\
$L_{\rm 3\,GHz}$ & $z$                &  0\% & 7\% &  4\% &  4\% & MAYBE \\
$L_{850\micron}$ & $L_{\rm 3\,GHz}$   & 53\% (34\%)& 21\% (33\%)& 33\% (48\%)& 46\% (58\%) & NO    (NO)    \\
$L_{850\micron}$ & $z$                & 53\% (34\%)& 88\% (75\%)& 20\% (34\%)& 35\% (52\%) & NO    (NO)    \\
$L_{850\micron}$ & $\alpha$           & 53\% (34\%)& 88\% (99\%)& 81\% (69\%)& 99\% (71\%) & NO    (NO)    \\
$L_{850\micron}$ & $D_{\rm lin}$      & 53\% (34\%)& 13\% ( 6\%)& 16\% ( 7\%)& 13\% ( 6\%) & NO    (NO)  \\
$L_{850\micron}$ & $L_{\rm \lya}$     & 59\% (41\%)& $-$ & 86\% (91\%)&  4\% ( 4\%) & MAYBE (MAYBE)  \\
$L_{850\micron}$ & Kmag               & 59\% (44\%)& $-$ & 47\% (33\%)& 88\% (72\%) & NO    (NO)    \\
$\alpha$    & $L_{\rm 3\,GHz}$        &  0\% &  56\% &  21\%&  11\% &    NO   \\
$\alpha$    &  $z$                    &  0\% &  2\% &  10\% &  8\%  & MAYBE     \\
\hline
\end{tabular}\\
$^{a}$ Cox's proportional hazard model only allows
censoring in the dependent variable, therefore this test could not be
applied with the \lya\ and $K$-band data.\\
\end{minipage}
\end{table*}

We have defined the subsample with redshift $z > 2.5$ to test for
selection effects and remove them from our sample.  The well known and
strong (see A01) radio power-redshift relation in flux limited samples
is likely to be the origin of many apparent correlations with redshift
(see Table \ref{SurvA_3}).
Another example of selection effects is the correlation found between
$L_{850}$ and radio spectral index $\alpha$ (Table \ref{SurvA_3}).  It
reflects the tight correlations between spectral index and redshift
(part of our search criteria for \hzrgs) and between $L_{850}$ and
redshift. In the $z > 2.5$ sample this correlation almost disappears
even though there is a large range in both $L_{850}$ and $\alpha$,
confirming that it was spurious. An additional advantage of this
subsample is the homogeneity of the supporting data (\lya\ fluxes and
$K$-band magnitudes are understandably sparse for lower redshift
sources).

Below, we describe the details of the survival analysis and the
criteria used to determine whether a correlation is present. In
Section \ref{correlations}, we discuss the correlations
individually.\\

\centerline{\it Survival analysis}

The application of survival analysis methods to astronomical data has
been described in detail by \citet{FeigelsonNelson85apj} and
\citet{Isobe86apj}. We have made use of the routines in the STSDAS
package of IRAF\footnote{IRAF is distributed by the National Optical
Astronomy Observatories, which are operated by the Association of
Universities for Research in Astronomy, Inc., under cooperative
agreement with the National Science Foundation.} \citep{Tody93adass},
which were modelled after the ASURV package
\citep{Lavalley92adass}. Cox's proportional hazard model, the
generalized Spearman's rank order correlation coefficient, and the
generalized Kendall's tau correlation coefficient test the null
hypothesis that no correlation is present in the sample. We adopt the
convention that two variables are correlated if the chance $P$ of the
null hypothesis being true is smaller than 5 per cent.  While these tests
should give similar results they have different specific limitations
and strong points. Therefore, a correlation is considered to be
reliable if all tests yield $P \la$ 5 per cent, and possible but
unconfirmed if only some indicate $P \la$ 5 per cent. Note that the Cox
test only allows censoring in the dependent variable and that the
generalized Spearman's Rho routine is not reliable for small data
sets ($N < 30$).  SCUBA occasionally yields negative flux
densities. We defined the $n$$\sigma$ upperlimit
for those cases to be $n \times $ the rms flux density. Similarly, the
$n\sigma$ upperlimit for a positive signal is defined as $S$ + $n
\times $ the rms flux density, with $S$ the observed signal. \\

\section{Correlations between parameters} \label{correlations}

Tables \ref{SurvA_3} and \ref{SurvA_hz3} represent the results of the
survival analysis as described above. We will now discuss the
motivation for each test and the implications of the correlations (or
lack thereof) found in order of importance.

\subsection{Redshift dependent submillimetre properties} 

\subsubsection{Flux density and relative detection fraction}

\begin{figure}
\epsfig{file=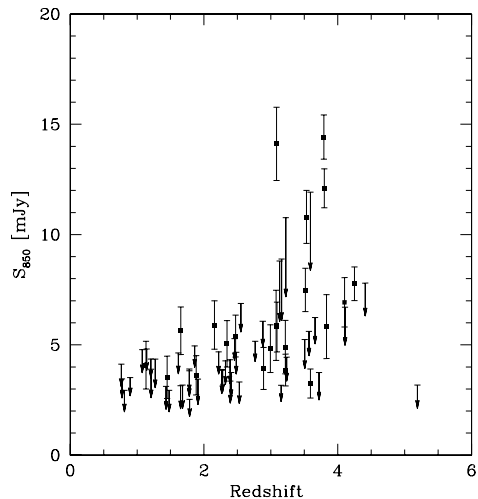,height=7cm,angle=0}
\caption{Observed 850\,\micron\ flux density and $3\sigma$ upper
limits versus redshift of all \nSampleAR\ radio galaxies discussed in this
paper. The size of the arrows and errorbars correspond to $1\sigma$
rms. \label{zS850}}
\end{figure}

\begin{figure}
\centering
\epsfig{file=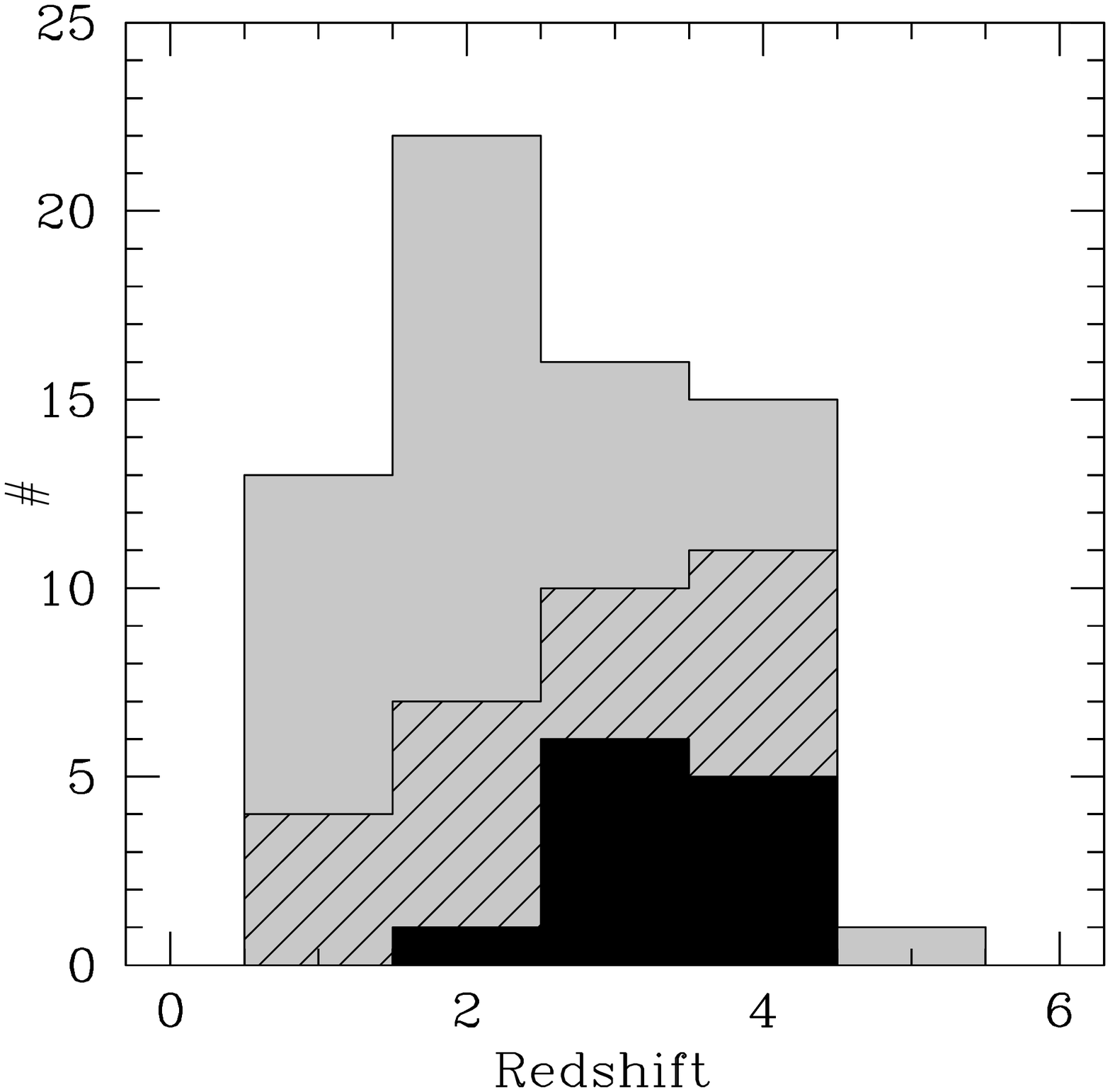,height=4cm,angle=0}
\epsfig{file=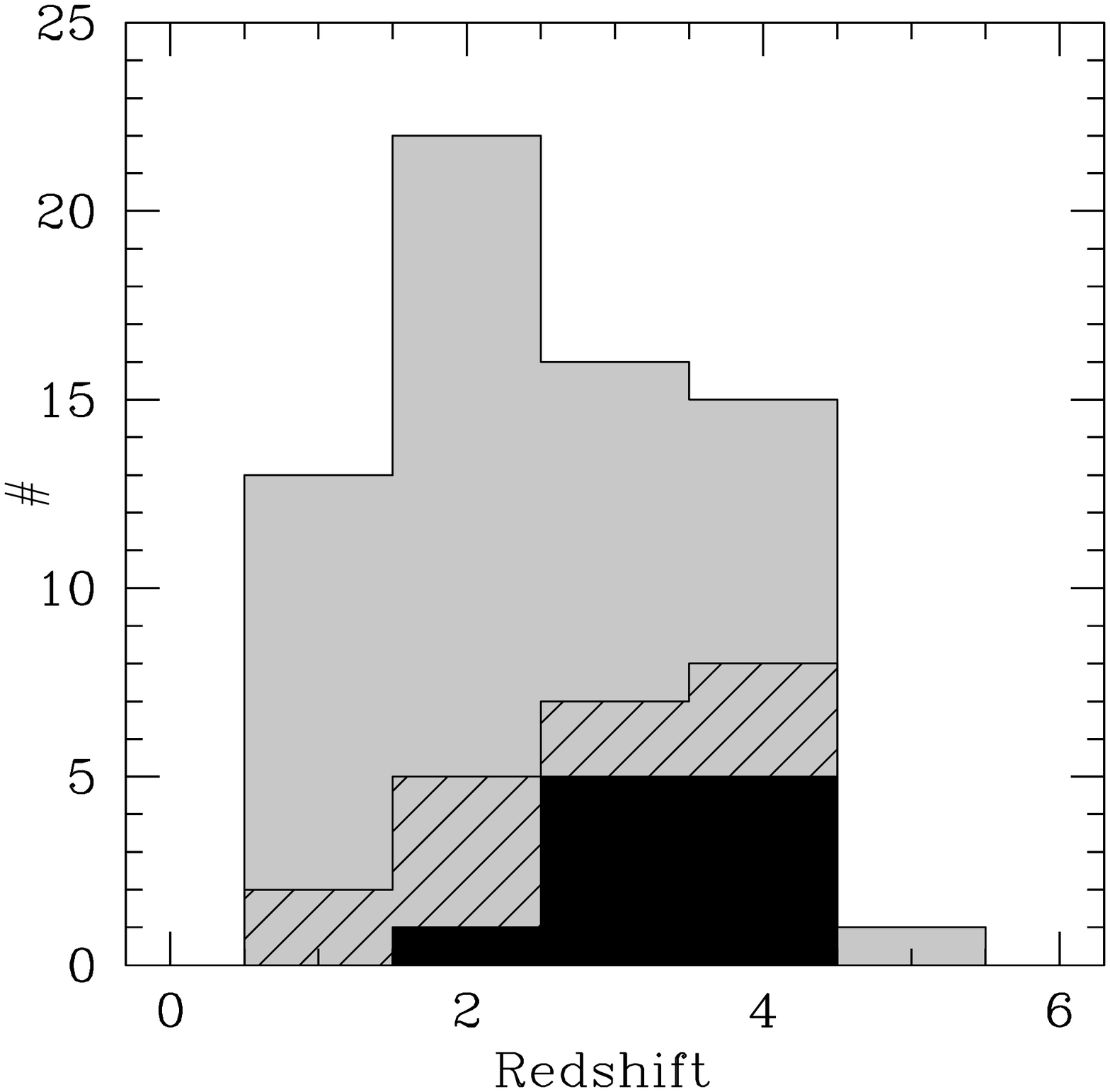,height=4cm,angle=0}
\caption{Histograms of 850\,\micron\ SCUBA detections with
S/N $>$ 2 (left) and S/N $>$ 3 (right) versus the total number of
observed radio galaxies (grey) as a function of redshift. The detections
from \citet{Archibald01mnras} are shown as dashed, the detections from
our program are shown in black.  At $z > 2.5$ $\sim$ 50--67 per cent of the
galaxies are detected, as opposed to $\sim$ 15 per cent at $z < 2.5$. The
bins have a width of unit redshift and are centered at $z =
1,2,3,4,5$. \label{DetHist}}
\end{figure}

Table \ref{SurvA_3} shows that the observed submm flux densities
$S_{850}$ are strongly correlated with redshift. This is reflected in
Figure \ref{zS850} and \ref{DetHist}. Figure \ref{zS850} shows the
observed 850\,\micron\ flux densities against redshift. Figure
\ref{DetHist} shows the number of radio galaxies that have been
observed in a particular redshift bin and the number of those which
have been detected given a $2\sigma$ or $3\sigma$ detection criterion.
The success rate of our $z > 3$ program is truly remarkable, and
corroborates the conclusion of A01 that the detection fraction of
$\sim$ 50--67 per cent at $z > 2.5$ is significantly different from
the detection fraction of $\sim$ 15 per cent at $z < 2.5$. Removing
the 4 brightest sources from the sample does not destroy this
relation.

\subsubsection{Investigation of the difference between detections and non-detections}

Given a 3$\sigma$ detection criterion the average submm flux density
of the 22 detected radio sources  and 45 non-detections are $<$$S_{850, \rm
\geq 3\sigma}$$>$ $= 6.60 \pm 0.71$\,mJy 
and $<$$S_{850, \rm <3\sigma}$$>$$ = 0.81 \pm 0.20$\,mJy respectively,
while the average for the entire sample is $<$$S_{850, \rm
sample}$$>$$ = 2.71 \pm 0.43$\,mJy. For a 2$\sigma$ detection
criterion the average submm flux density of the 30 detected galaxies
and 37 non-detections are $<$$S_{850, \rm \geq 2\sigma}$$>$$ = 5.29
\pm 0.60$\,mJy and $<$$S_{850, \rm <2\sigma}$$>$$ = 0.34 \pm
0.19$\,mJy respectively.

\begin{figure}
\epsfig{file=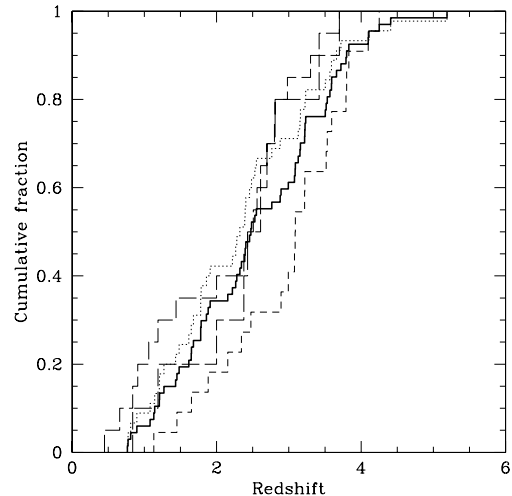,height=7cm,angle=0}
\caption{Cumulative histograms $\Sigma N(z)$ for redshifts of submm
sources from the literature and the \nSampleAR\ radio galaxies (solid line)
discussed in this paper. The short-dashed and the dotted line indicate
submm detections of radio galaxies versus non-detections given a 3$\sigma$
detection criterion, respectively. The detections have a higher median
redshift ($z = 3.1$) than the parent sample ($z = 2.5$).
The two long-dashed lines represent $\Sigma N(z)$ of the 10 submm
sources with spectroscopic redshifts found by \citet{Chapman03nat} and
using 9 additional spectroscopic redshifts
\citep{Ivison98mnras,Ivison00apj,Dey99apj,Soucail99aa,Eales00aj,Ledlow02apj,Chapman02bmnras,Chapman02apj,Aretxaga03mnras,Frayer03aj}.\label{CumFracHist}}
\end{figure}

Figure \ref{CumFracHist} shows cumulative redshift distributions
$\Sigma(z)$ for various subsets of the radio galaxy sample together
with spectroscopic redshifts for other submm samples from the
literature.  As was suspected from Figure \ref{DetHist}, the radio
galaxies detected in the submm follow a distribution that differs
significantly from both the undetected sources and the parent
sample. The K--S test shows with $P >$ 99 per cent confidence that
using a 2$\sigma$ or 3$\sigma$ detection criterion picks out the same
population (both for the detections and for the non-detections),
whereas the chance that the detections and non-detections are
distributed similarly is $P <$ 1 per cent.

Does the high median redshift ($z= 3.1$) of the detected sources (the
parent sample has $z = 2.5$) purely reflect the strong negative
$K$-correction or does it reflect a change in \LFIR?  Figure
\ref{CumFracHist} shows that the median redshift of the detections is
higher than the median redshift ($z = 2.4$) for SCUBA sources
\citep{Chapman03nat}, while the redshift distribution of the parent
sample follows the SCUBA population closely. This argues in favor of a
different \LFIR\ for detections and non-detections.  What then causes
the difference in redshift distribution between detected \hzrgs\ and
the submm population? There are at least two possible explanations. It
may indicate that \hzrgs\ are more massive, at the centers of
protoclusters \citep{Venemans03confproc}, and therefore undergo a
faster evolution than `normal' submm sources and finish the bulk of
their formation process early. Alternatively, it may reflect a
selection effect as the requirement of a faint radio counterpart prior
to spectroscopic follow-up for the submm sample selects against high
redshift galaxies \citep{Chapman03nat}.

\subsubsection{The increase of submm luminosity with redshift} \label{Lumdecline}

\begin{figure}
\epsfig{file=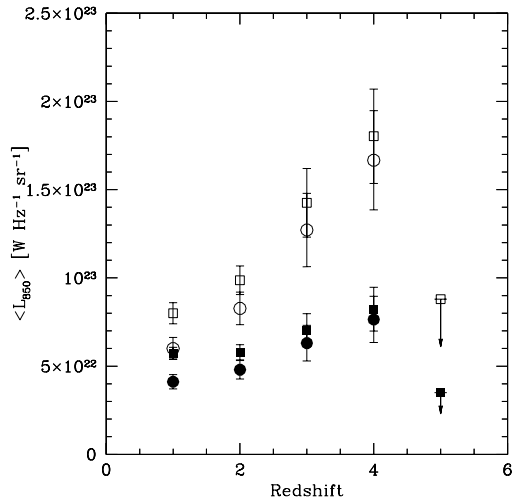,height=7cm,angle=0}
\caption{The average rest-frame 850 \micron\ luminosity versus
redshift for \nSampleAR\ radio galaxies as estimated using the
Kaplan--Maier estimator for each redshift bin. The data points
indicated with open squares and circles were calculated assuming the
same dust template as A01 ($\beta=1.5$, $T_{\rm d} = 40$\,K) for
3$\sigma$ and 2$\sigma$ detection criteria respectively. The solid
symbols are similar but assume $\beta=2.0$ instead.  The data points
at $z = 5$ are $3\sigma$ upperlimits based on the non-detection of
TN~J0924$-$2201.
\label{L850z}}
\end{figure}

Figure \ref{L850z} shows the inferred 850\,\micron\ luminosity
$L_{850}$ as a function of redshift.  For each redshift bin $L_{850}$
has been estimated using the Kaplan--Maier estimator, a survival
analysis technique which tries to estimate the true distribution of
the underlying population by incorporating both upperlimits and
detections. We have computed the luminosities both using a dust
template with $\beta = 1.5$ and $\beta = 2.0$. 
Adding our data to the sample of A01 confirms that \hzrgs\ have higher
submm luminosities than lower redshift sources.  However, there are
significant differences with the findings from A01: (i) $L_{850}$ at
low redshifts is higher than inferred by A01. This is because we have
chosen, not to correct for possible synchrotron contamination (see
Section 2.3) (ii) There is weak evidence for a turnover or leveling
off at $z > 4$ (based on the lower flux densities of the five galaxies
with $z> 4$ compared to redshifts $3 < z < 4$; see
Fig. \ref{zS850}). This explains why there is no strong correlation
between $L_{850}$ and $z$ in the high redshift subsample (see Table
\ref{SurvA_hz3}). (iii) A01 remarked that the increase in $L_{850}$
becomes less pronounced if one assumes a dust template with $\beta =
2$. Because $L_{850}$ at low redshifts is higher than in A01, this
effect is even stronger in our sample and we find that $L_{850}$ may
be rather constant. The sources would still be highly submm luminous,
implying huge dust masses and star formation rates.

\subsection{The connection between submm and radio luminosity}

Radio galaxies host luminous AGNs and for favorable geometries (\eg\
dusty warped disks) heating of dust by only a small fraction ($\leq
20$ per cent) of their UV radiation could easily explain typical far-IR
luminosities \citep[\eg\ ][]{Sanders89apj}. Therefore, an important
question is: do the AGN heat the large-scale dust significantly,
i.e. are $L_{\rm radio}$ and \LFIR\ correlated?

To answer this question we have estimated the rest-frame radio power
at 3\,GHz, $L_{3\rm GHz}$. We have chosen this frequency because for
the redshift range $1.1<z<7.2$ this requires only interpolation
between between 365\,MHz and 1.4\,Ghz.
Of course this assumes that the UV emission of the central source
relates to its radio output \citep[\eg ][]{Willott99mnras,DeBreuck00aa}.

\begin{figure}
\epsfig{file=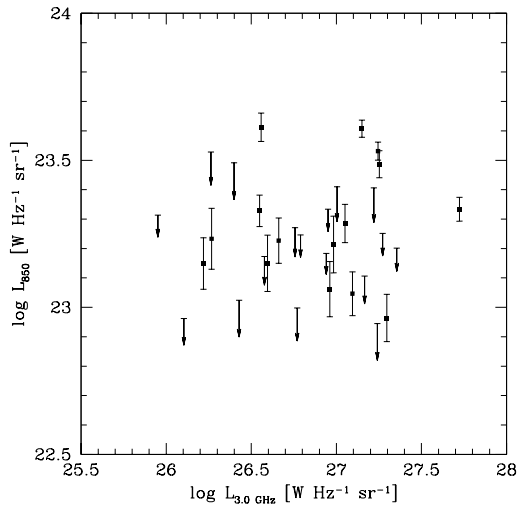,height=7cm,angle=0}
\caption{Submillimetre luminosity against radio power $L_{\rm 3 GHz}$
for 32 radio galaxies with $z > 2.5$.  No strong correlation is
apparent, indicating that heating of the dust by AGN is not a dominant
process for radio galaxies. Symbols as in Figure \ref{zS850}.
\label{Lumradio}}
\end{figure}

While Table \ref{SurvA_3} shows that there is a possible correlation
between $L_{850}$ and $L_{3\rm GHz}$ over the entire redshift range of
our sample, this is probably an effect of the strong redshift
dependence of both $L_{850}$ and $L_{3\rm GHz}$. The tests represented
in Table \ref{SurvA_hz3} only take radio galaxies with $z > 2.5$ into
account and even though this subsample contains almost all submm
detections the correlation disappears. The scatterplot in Figure
\ref{Lumradio} is a graphical representation of the same result. 
We don't find evidence for dust heating by AGN,
in accordance with earlier findings \citep[\eg\
A01;][]{Willott02mnras,Andreani02aa} but see \citet{Haas03aa} for a
contrasting view based on ISO observations of local (most have $0.1 <
z < 1.0$) hyperluminous quasars.  This analysis is likely to be an
oversimplification because the typical time-scales for starburst and
radio activity differ by an order of magnitude.

\subsection{An anti-correlation between submm flux and UV polarisation}

Optical polarimetry studies can be used to search for direct signs of
starburst activity and determine the relative contributions of AGN and
starburst to the UV continuum \citep[\eg ][]{Dey97apj,Vernet01aaa}.
If the starburst dominates the UV/optical light, then the scattered
(i.e. polarized) AGN light is diluted and one expects a low
polarisation fraction of the UV continuum and perhaps a correlation
with the observed submm flux densities.

\begin{figure}
\epsfig{file=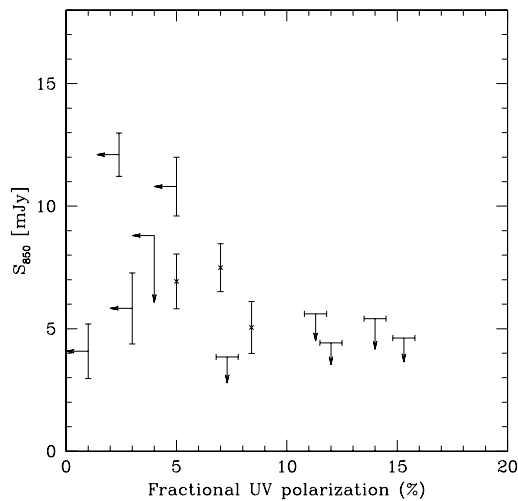,height=7cm,angle=0}
\caption{850\,\micron\ flux $S_{850}$ against observed UV continuum
polarisation fraction. None of the highly polarized sources are
detected in the submm, indicating that low polarisation (indicative
of a starburst) and high \LFIR\ both trace star formation.
\label{UVpolS850}}
\end{figure}

Polarimetry results exist for 13 of the galaxies in our sample
\citep[][De Breuck \etal\ in preparation]{Dey97apj,Vernet01aaa}.  A
summary of the polarisation fractions and corresponding 850\,\micron\
flux densities is given in Table \ref{PolTable}.  We have plotted the
observed submm flux density against UV continuum polarisation
fraction in Figure \ref{UVpolS850}. No sources with high UV
polarisation are detected in the submm. This is supported formally by
survival analysis: Kendall Tau gives a probability $P = 3$ per cent that
there is no correlation.

\begin{table}
\centering
\begin{minipage}{8cm}
\caption{UV/optical polarisation fractions and submm
fluxes for all radio galaxies for which both have been
observed. UV/optical polarisation data are taken from
\citet{Dey97apj}, \citet{Vernet01aaa}, and De Breuck \etal\ in preparation, the
submm data are from A01 and this paper. \label{PolTable}}
\begin{tabular}{lrrr}
\hline
Source & $z$ & P(\%) & $S_{850}$ (mJy) \\ 
\hline
3C356          & 1.079  &    14  &  $<4.8$   \\
3C368          & 1.132  &  $<1$  &  4.1      \\
3C324          & 1.206  &    12  &  $<4.4$   \\
4C40.36        & 2.265  &   7.3  &  $<3.9$   \\
4C48.48        & 2.343  &   8.4  &  5.1      \\
4C23.56        & 2.483  &  15.3  &  $<4.7$   \\
MRC0316$-$257  & 3.130  &  $<4$  &  $<8.9$   \\
TNJ0121+1320   & 3.516  &     7  &  7.5      \\
TX1243+036     & 3.570  &  11.3  &  $<5.6$   \\
6C1908+722     & 3.532  &  $<5$  & 10.8      \\ 
4C41.17        & 3.798  & $<2.4$ & 12.1      \\
TNJ2007$-$1316 & 3.830  &  $<3$  &  5.8      \\
TNJ1338$-$1942 & 4.100  &     5  &  6.9      \\
\hline
\end{tabular}
\end{minipage}\end{table}

This result is consistent with the view that for \hzrgs\ star
formation and submm emission are closely linked while any AGN
contribution to the FIR is negligible. \citet{Tadhunter02mnras} also
found some evidence for this. Two out of the three objects which are
starburst dominated in their sample of 2-Jy radio galaxies are also
extremely FIR luminous. Further support for low polarisation in
starbursting systems comes from the luminous submm source
SMM~J02399$-$0136 \citep{Vernet01aab} which shows only moderate ($P
\sim$ 5 per cent) polarisation.

The weakness of the submm luminosity in highly polarized sources
together with the tentative correlation between \lya\ and submm flux
(see next section) is consistent with the anti-correlation between
\lya\ luminosity and UV polarisation reported by \citet{Vernet01aaa}.
They argue that \lya\ photons are resonantly destructed by dust and
that scattering by the same dust results in a higher fractional
polarisation. This explanation, however, seems at odds with the lower
submm flux densities in highly polarized sources unless there would
be significant amounts of dust that is too hot to be detected in the
submm.

\subsection{Submillimetre and Ly$\alpha$ flux}

Since many \hzrgs\ are embedded in giant \lya\ haloes \citep[\eg\
][]{McCarthy93araa,vanOjik96aa,Reuland03apj}
it would be interesting to see if one can relate the amount of
available star formation `fuel', as estimated roughly by the spatial
size, or luminosity, of these emission line haloes and the amount of HI
as probed by \lya\ absorption to the observed rest-frame FIR
luminosities. Alternatively, one might expect a strong
anti-correlation due to destruction of \lya\ emission by dust.

\begin{figure*}
\centering
\hspace{-0.75cm}
\epsfig{file=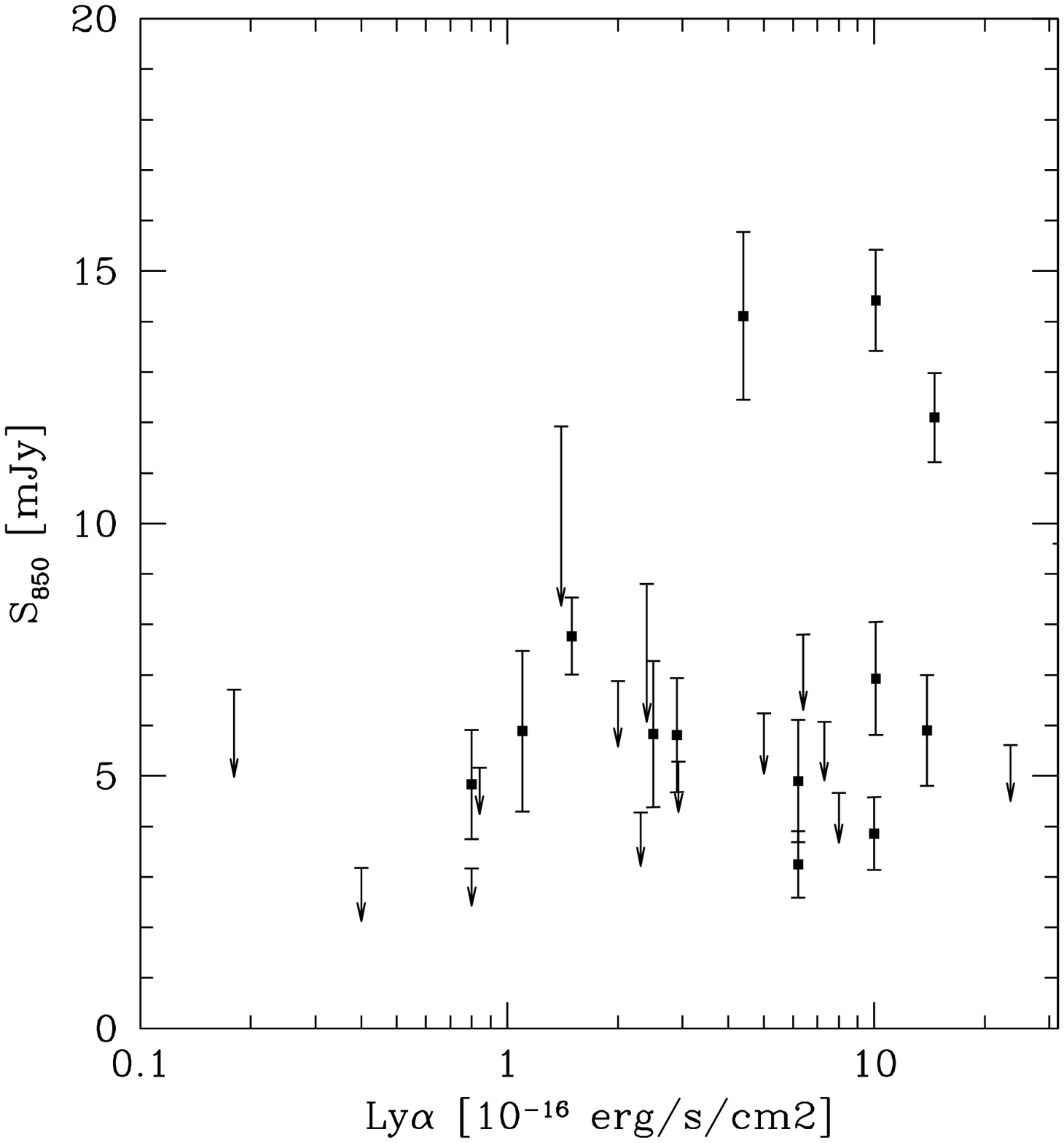,height=7cm,angle=0}
\hspace{1.40cm}
\epsfig{file=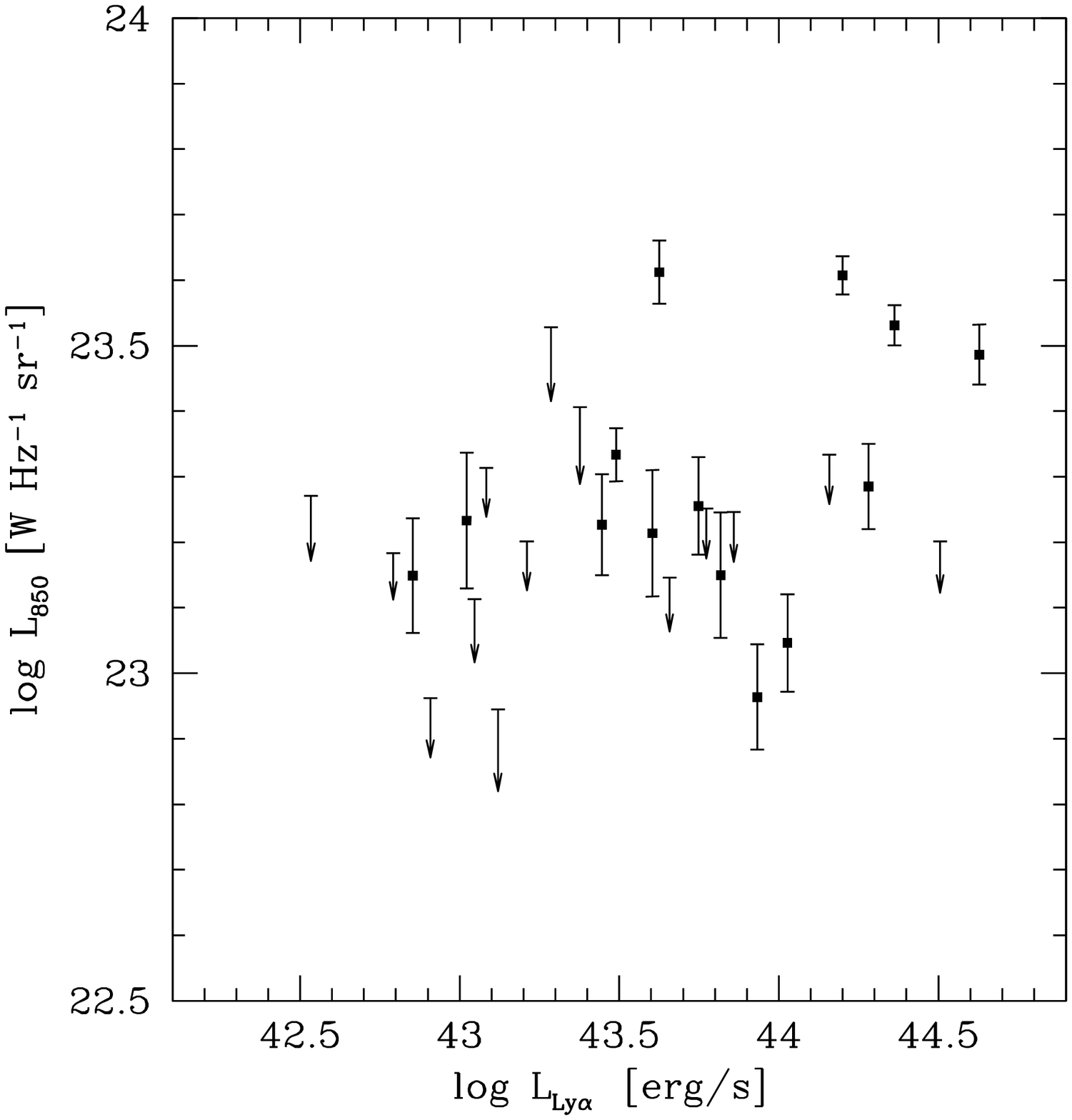,height=7cm,angle=0}
\caption{$S_{850\,\micron}$ against observed \lya\ line flux (left)
and inferred luminosities $L_{850}$ against $L_{\lya}$ (right).
Tentative evidence for a correlation is apparent.  The galaxies
brightest in \lya\ seem to have higher submm flux densities, possibly
indicating that galaxies with richer gaseous environments (as traced
by \lya) also host more massive starbursts.  Symbols as in Figure
\ref{zS850}.
\label{SLyaLum}}
\end{figure*}

Many \hzrgs\ are identified based on their \lya\ line and good data on
the observed \lya\ fluxes is available in most cases.  There is a
strong selection effect, however, since sources with low \lya\ flux densities
are less likely to be recognized as \hzrgs.  Interestingly, Table
\ref{SurvA_hz3} and Figure \ref{SLyaLum} indicate that a correlation
may be present. While this relation is tentative only, it offers
perspectives for future programs.  Better data should make it possible
to investigate whether there is a possible relationship between the
dust content and \lya\ to \civ\ or \nv\ emission line
ratio, as indicators of metalicity \citep[c.f. ][]{Vernet01aaa}. This
would be especially interesting in the light of findings by
\citet{HamannFerland99araa} and \citet{Fan01aj} that QSOs show
(super)solar abundances out to the highest redshifts and must already
have undergone significant star formation. A comparison with the more
readily studied \hzrg\ host galaxies would be crucial to better
understand this result.

The detection of both \lya\ and dust in many \hzrgs\ is intriguing
because already a small amount of dust can extinguish \lya\ radiation
efficiently. The detection of \lya\ in the submm sources of
\citet{Chapman03nat} is equally surprising. One explanation may be
that \lya\ emission and dust are located predominantly in different
locations (\eg\ a merger between one dusty and one less dusty
component). Some evidence for this comes from 4C~60.07 at $z = 3.8$ in
which the spatially resolved dust continuum emission
\citep{Papadopoulos00apj} is anti-correlated with the \lya\ emission
\citep[Fig. 5 in][]{Reuland03apj}. \citet{Smail03mnras} also report
evidence for an offset between far-IR and UV emitting regions in the
$z=2.38$ starburst galaxy N2~850.4.

\subsection{$L_{850}$ and linear size}

\begin{figure}
\centering
\epsfig{file=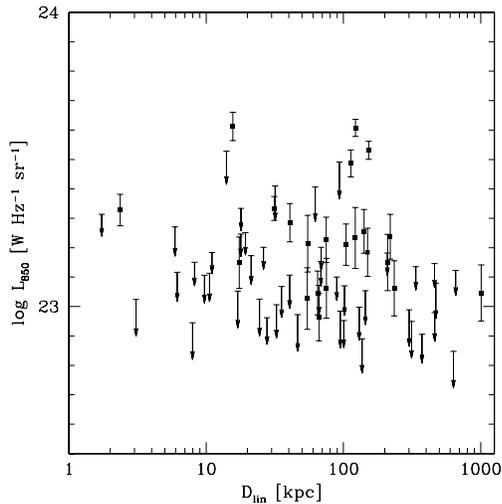,height=7cm,angle=0}
\caption{850 \micron\ luminosity $L_{850}$ against the radio
source projected linear size $D_{\rm lin}$. There is no evidence for
a strong anti-correlation as was found for radio loud quasars by
\citet{Willott02mnras}. Symbols as in Figure \ref{zS850}.
\label{DlinLum}}
\end{figure}

\citet{Willott02mnras} found an anti-correlation between linear size
and 850\,\micron\ luminosity for their sample of $z \sim 1.5$ radio
loud quasars.  This effect was attributed to a possible relation
between the jet-triggering event and a short-lived starburst or
quasar-heated dust in QSOs. No such correlation was found for a
matched subset of the A01 sample centered at the same redshift. We
have briefly investigated whether such a correlation exists in the
present sample. Figure \ref{DlinLum} shows the observed submm flux
plotted against the projected linear size.  No correction for possible
differences in inclination angle was made because such corrections are
expected to be relatively small among radio galaxies \citep[even when
comparing radio galaxies with quasars projection effects result in
only a factor $\approx 1.6$ difference in projected linear size for
$\theta_{\rm trans} =
53\degree$;][]{Willott00mnras,Willott02mnras}. Our analysis does not
support such a strong relation and if one interprets linear size as an
estimate of the age of the radio source \citep[reasonable if one
assumes that the environments and jet powers do not change
significantly; see \eg ][]{Blundell99aj}, then this would be
consistent with the scenario sketched by Willott \etal\ that the
correlation found for QSOs is indicative of short starbursts while the
starbursts in \hzrgs\ have longer time-scales and could be forming the
bulk of the stellar populations. Despite the absence of a clear
correlation, some relation between age of radio source and starburst
might have been expected if the black-hole and stellar bulge grow in a
symbiotic fashion \citep[\eg ][]{Williams99mnras}. However, in
realistic scenarios many complicating factors can be envisaged such as
the significantly different time-scales involved. Growth of the radio
source may first induce star formation by compressing molecular clouds
\citep{BegelmanCioffi89apj,Bicknell00apj} and may then actually signal
the end of the starburst by clearing out all the cool gas from the
central region \citep{Rawlings03confproc}.

\subsection{Submillimetre and near-IR emission}

\begin{figure}
\epsfig{file=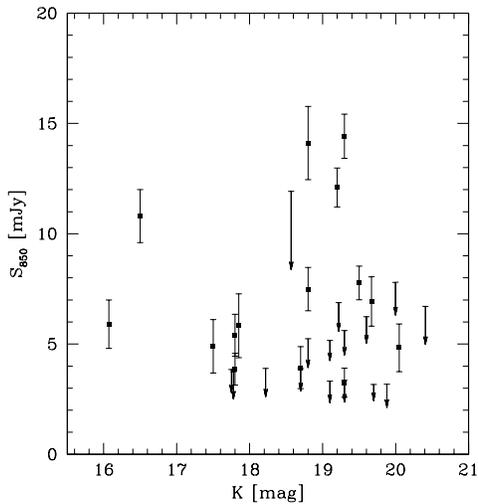,height=7cm,angle=0}
\caption{$S_{850\,\micron}$ against
observed $K$-band magnitude.  No correlation is apparent, and no
relation between \LFIR\ and stellar mass/optical star formation rate of the
host galaxy can be established. Symbols as in Figure \ref{zS850}.
\label{KS850}}
\end{figure}

As discussed by \citet{Isaak02mnras}, for QSOs one might naively
expect a correlation between the submm and optical flux
densities. This correlation might arise regardless of whether stars or
a buried AGN are responsible for heating of the dust since black hole
mass scales with stellar bulge mass. \citet{Priddey03mnras} note that
the picture is likely to be more complex and that any correlation
might be smeared out due to differences in relative timing between AGN
fueling and starburst, or varying dust-torus geometries.  Radio
galaxies might be better suited for such studies, since according to
orientation based unification schemes \citep[\eg][]{Barthel89apj} the
AGN are obscured by a natural coronograph and rest-frame $B$-band
luminosity therefore is a fairly reliable measure of the mass of the
stellar population or starburst activity.  For this reason we have
correlated the $K$-band magnitudes (samples rest-frame $B$,$V$ for $z
> 3$) with the submm flux densities. As can be seen from Table
\ref{SurvA_hz3} and Figure \ref{KS850} no such correlation is
apparent. This remains true, if we apply a $K$-correction to the
near-IR magnitudes. However, the dataset is rather inhomogeneous and
more sensitive data in both the submm and near-IR are required to
truly rule out any correlation. Sensitive multi-band observations
could be used to (i) construct a redshift independent estimate of the
line-free rest-frame optical continuum for comparison with \LFIR\
\citep[\protect{[\oiii]} and H$\alpha$ can sometimes dominate the
$K$-band \eg\ 4C~30.36, 4C~39.37;][]{Egami03aj}, and (ii) check
whether FIR emission is stronger for sources with higher intrinsic
reddening \citep[following
\eg][]{Calzetti97aj,AdelbergerSteidel00apj,Seibert02aj}.

\begin{figure*}
\centering
\epsfig{file=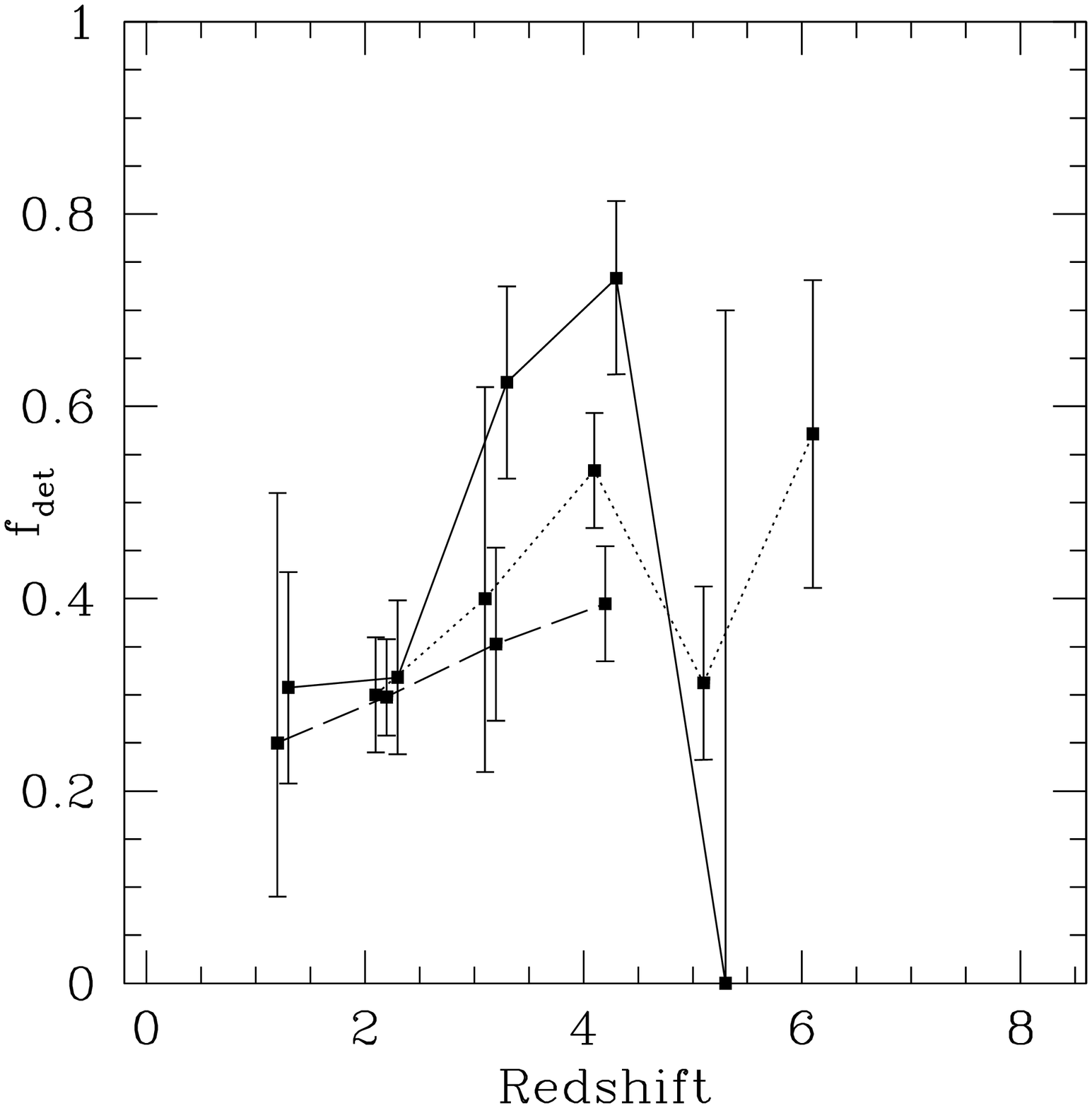,height=7cm,angle=0}
\hspace{1cm}
\epsfig{file=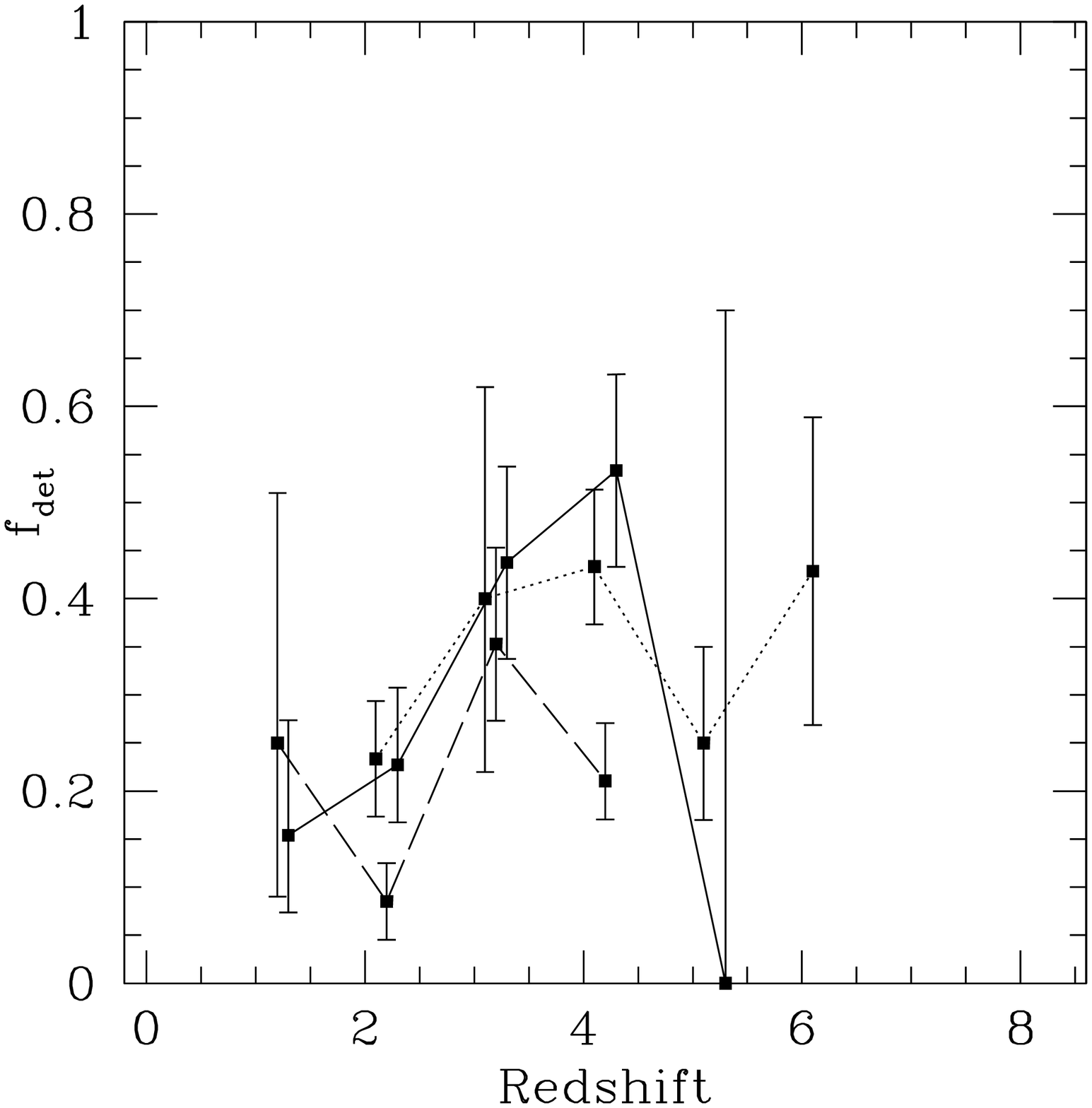,height=7cm,angle=0}
\caption{The 850\,\micron\ detection (left $S/N > 2$; right $S/N > 3$)
fractions of radio galaxies as function of redshift (solid line).  The
fractions rise up to a redshift of $z = 4$. For comparison we also
plot the detection fraction for observations of radio quiet QSOs at $z
=$ 1 -- 5, and radio loud quasars at $z \sim 1.5$ (after matching the
sensitivity to 3\,mJy rms) at 850\,\micron\ \citep[dashed
line;][]{Priddey03mnras,Isaak02mnras,Willott02mnras} and 1.25\,mm
\citep[dotted
line;][]{Omont01aa,Omont03aa,Carilli01apj,Petric03aj,Bertoldi03aa}. The
detection rates for the QSOs appear to follow the same trend as for
the radio galaxies, but the QSO datapoint at $z = 5$ seems indicative
of a turnover (the $z = 5$ point for the \hzrgs\ is based on only 1
galaxy, but taking all $z>4$ \hzrgs\ yields a similar result). The
datapoint at $z = 6$ could indicate a detection fraction larger even
than what is observed for $z = 4$. This might reflect, the extreme
properties of those sources.
\label{DetFrac}}
\end{figure*}

\subsection{Summary}

From our analysis we consider the following results the most interesting:
\begin{enumerate}
\item $S_{850}$ and $z$ are strongly correlated. Whether this is true
also for \LFIR\ and $z$ depends on the assumed dust template.
\item $L_{850}$ and radio power $L_{3 \rm GHz}$ do not correlate strongly,
indicating that AGN do not dominate the submm emission (either
through synchrotron contamination or dust heating by the UV
continuum).
\item $S_{850}$ and \lya\ appear weakly correlated. While this is not
very convincing, the interesting result is the absence of a strong
anti-correlation as expected naively from the destruction of \lya\
emission by dust.
\item $S_{850}$ and UV polarisation fraction appear
anti-correlated. This is expected if the UV continuum of the buried
AGN does not contribute significantly to the dust heating.
\end{enumerate}

\section{A comparison of radio galaxies with QSOs}

The optically thin (sub-)mm emission of starbursts should be largely
independent of viewing angle.  Orientation based unification schemes
for AGN \citep[\eg][]{Barthel89apj} therefore suggest that we can
combine observations of radio galaxies with those of QSOs, allowing us
to study the evolution of star formation in their host galaxies over a
larger redshift range. Extensive surveys of high-redshift (mostly
radio-quiet) QSOs have been published recently
\citep{Carilli01apj,Omont01aa,Omont03aa,Isaak02mnras,Priddey03mnras,Petric03aj,Bertoldi03aa}.
We have combined their results and compare those with our sample.

\begin{figure}
\epsfig{file=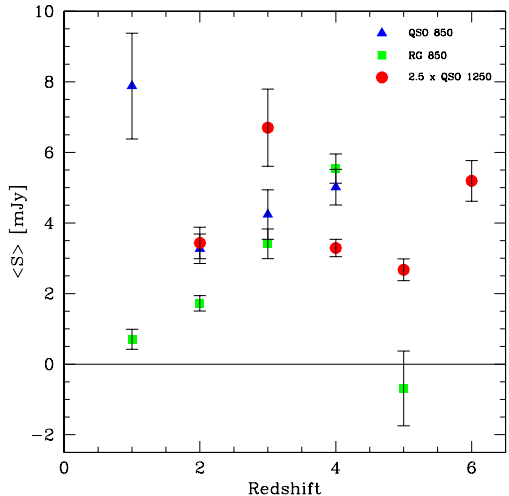,height=7cm,angle=0}
\caption{The average measured (sub-)mm flux density binned in redshift
has been plotted for QSOs and radio galaxies.  The squares represent
observations of radio galaxies at 850\,\micron, the triangles QSOs at
850\,\micron, while the circles represent observations of QSOs at
1250\,\micron\ multiplied by a factor of 2.5 for easier
comparison. The datapoint at $z \approx 1.5$ represents the radio loud
quasars from \citet{Willott02mnras}. The trends with redshift are
similar to what is seen in Figure \ref{DetFrac} and confirm the steady
rise from $z =$ 1 to $z = 4$ for the galaxies.
\label{avgS850}}
\end{figure}

Figure \ref{DetFrac} shows the detection fraction as a function of
redshift for the QSO and radio galaxy samples.  In many redshift bins
only a small number of detections and non-detections is responsible
for estimating the success rates for the present effective flux
limits.  We have estimated the range over which the `true success
rate' could vary such that there would be a 68 per cent chance of
measuring the observed detection fraction. This range is indicated by
the errorbars.  Figure \ref{avgS850} shows the average \mbox{(sub-)}mm
flux density for the various samples as a function of redshift.

Reassuringly, the trends for the QSOs and radio galaxies shown in
these figures mimic each other.  The average observed \mbox{(sub-)}mm
flux density rises to redshifts of $z\sim4$.  For higher redshifts
there is some evidence for a decline (see the discussion in Section
\ref{Lumdecline} and the z = 5 data point of the MAMBO QSO
obervations). However, the observations of $z > 4$ QSOs would also be
consistent with a fairly constant or even a rise in the detection
fraction and average FIR luminosity out to the highest redshifts. This
depends rather critically on the $z \sim 6$ QSOs, which are likely to
be atypical even for QSOs.

The average flux density and detection rate of the QSOs is comparable
to that of the \hzrgs, even though the \hzrg\ surveys are more
sensitive than the QSO surveys by a factor of approximately 3. This,
coupled with the assumption that quasars are on average more luminous
than radio galaxies in AGN unification schemes based on receding torus
models \citep[\eg ][]{Simpson03confproc}, may be viewed as further
support for an AGN related component in QSOs, similar to what was
found for radio quasars \citep{Andreani02aa,Willott02mnras}.
Note that the quasars discussed by \citet{Willott02mnras} are
exceptionally bright compared to the mostly radio quiet QSOs
represented in Figs. \ref{DetFrac},\ref{avgS850}.

Due to a lack of carefully matched surveys fundamental differences
between type I and type II AGN and radio-loud versus radio-quiet
targets cannot be properly investigated yet, although slowly
progress is being made.

\section{Summary}

We have presented SCUBA observations of 24 radio galaxies and compared
those with earlier results of 47 radio galaxies from the survey by
A01.  We confirm that \hzrgs\ are massive forming galaxies, forming
stars up to rates of a few thousand \Msunpyr\ and that there is no
strong evidence for a correlation with radio power. Further evidence
for a predominantly starburst nature of the far-IR emission comes from
the striking anti-correlation between submm flux density and UV
polarisation (Fig. \ref{UVpolS850}).

In agreement with A01 we find that submm detection rate appears to be
primarily a function of redshift. If this is interpreted as being due
to a change in the intrinsic far-IR luminosity, it would be consistent
with a scenario in which the bulk of the stellar population of radio
galaxies forms rapidly around redshifts of $z = 3-5$ after which they
are more passively evolving \citep[c.f. ][]{Best98mnras}. We also find
that the median redshift of the \hzrgs\ with SCUBA detections ($z =
3.1$) is higher than the median redshift of the submm population
\citep[$z = 2.4$;][]{Chapman03nat}. In the current picture of
hierarchical galaxy formation, this could be interpreted as that
\hzrgs\ are more massive galaxies, which are then thought to begin
their collapse at earlier cosmic times and evolve faster and finish
the bulk of their formation process earlier. Alternatively, it could
indicate that higher redshift submm sources are being missed due to
the requirement of a radio counterpart prior to spectroscopic
follow-up.

\hzrgs\ have accurately determined redshifts and host identifications
and are thought to be the most massive galaxies at any epoch
\citep{DeBreuck02aj}.  Therefore, \hzrgs\ are a key population for
studies of galaxy formation in the early universe, allowing detailed
follow-up mm-interferometry observations to study their dust and gas
content. Currently, they offer the best way to obtain reliable dynamic
masses for a significant number of massive high redshift galaxies.  These
\hzrgs\ would be especially suited to constrain any evolution in
galaxy mass with redshift, study changes in evolutionary status, gas
mass, and the starburst-AGN connection. If all of these turn out to
have masses larger than $10^{11}$\Msun, then this could have important
consequences (depending on rather uncertain correction factors for the
fraction of similar galaxies for which the black hole is dormant) for
our understanding of galaxy formation, because only few such massive
galaxies are expected at such high redshifts \citep{Genzel03apj}.

\section*{Acknowledgments}

We thank Chris Willott and Steve Rawlings for useful discussions and
the anonymous referee for helpful suggestions to improve the paper.
We gratefully acknowledge the excellent help of the JCMT staff and
`flexible' observers who collected data for our program. In particular
the help of Remo Tilanus was indispensable.  The authors wish to
extend special thanks to those of Hawaiian ancestry on whose sacred
mountain we are priviledged to be guests. The JCMT is operated by JAC,
Hilo, on behalf of the parent organizations of the Particle Physics
and Astronomy Research Council in the UK, the National Research
Council in Canada and the Scientific Research Organization of the
Netherlands.  The work of M.R. and W.v.B.  was performed under the
auspices of the U.S. Department of Energy, National Nuclear Security
Administration by the University of California, Lawrence Livermore
National Laboratory under contract No. W-7405-Eng-48.


\bsp 

\label{lastpage}

\end{document}